\documentclass[conference]{IEEEtran}
\IEEEoverridecommandlockouts
\usepackage{cite}
\usepackage{amsmath,amssymb,amsfonts}
\usepackage{algorithmic}
\usepackage{graphicx}
\usepackage{textcomp}
\usepackage{xcolor}
\newcommand{\qq}[1]{``#1''}

\usepackage{dsfont} 
\usepackage[skip=2pt,font=small]{caption}
\usepackage{subcaption}
\usepackage{array}
\newcolumntype{L}{>{\centering\arraybackslash}m{1.7cm}}
\newcolumntype{M}{>{\centering\arraybackslash}m{1.7cm}}
\newcolumntype{N}{>{\centering\arraybackslash}m{3.808cm}}
\def\BibTeX{{\rm B\kern-.05em{\sc i\kern-.025em b}\kern-.08em
    T\kern-.1667em\lower.7ex\hbox{E}\kern-.125emX}}
    
\newtheorem{theorem}{Theorem}
\newtheorem{lemma}{Lemma}
\newtheorem{proposition}[theorem]{Proposition}

\newtheorem{conjecture}[theorem]{Conjecture}

\newcommand{\R}{\mathbb{R}}
\newcommand{\pr}{\mathbb{P}}
\newcommand{\E}{\mathbb{E}}
\newcommand{\N}{\mathbb{N}_0}

\newcommand{\G}{\Gamma}

\newcommand{\event}{\mathcal{E}}
\newcommand{\limit}{\underset{n \rightarrow \infty}\lim}
\newcommand{\hhh}{\mathbb{H}(n;K)} 
\newcommand{\hh}{\mathbb{H}(n;\mu,K_n)} 
\newcommand{\nodes}{\mathcal{N}}
\newcommand{\kk}{\langle K_{n} \rangle}
\newcommand{\K}{ K_{n}}

\newcommand{\dd}{{\rm deg}} 
\newcommand{\ii}{\mathds{1}} 
\newcommand{\xd}{Z_d(n;\mu,K_n)}
\newcommand{\zl}{X_{k-1}(n;\mu,K_n)}
\newcommand{\g}{\gamma}
\newcommand{\con}{ \rightarrow}
\newcommand{\dcon}{\not \rightarrow}

\def \OO {\mathrm{O}}
\def \oo {\mathrm{o}}

\newcommand{\fsquare}{\vrule height6pt width7pt depth1pt}
\newcommand{\myendpf}{\hfill\fsquare \\[0.1in]}

\usepackage{mathtools}

\begin{document}

\title{Towards $k$-connectivity in Heterogeneous Sensor Networks under Pairwise Key Predistribution 
\thanks{This work was supported in part by the National Science Foundation through grant CCF   
\#1617934.}
}

\author{\IEEEauthorblockN{Mansi Sood and Osman Ya\u{g}an}
\IEEEauthorblockA{Department
of Electrical and Computer Engineering and CyLab, \\
Carnegie Mellon University, Pittsburgh,
PA, 15213 USA\\
msood@cmu.edu, oyagan@ece.cmu.edu}}

\maketitle

\begin{abstract}
We study the secure and reliable connectivity of  wireless sensor networks under the {\em heterogeneous} pairwise key predistribution scheme. This scheme was recently introduced as an extension of the  random pairwise key predistribution scheme of Chan et al. to accommodate networks where the constituent sensors have different capabilities or  requirements for security and connectivity. For simplicity, we consider a heterogeneous network where each of the $n$ sensors is classified as type-1 (respectively, type-2) with probability $\mu$ (respectively, $1-\mu)$ where $0<\mu<1$. Each type-1 (respectively, type-2) node selects 1 (respectively, $K_n$) other nodes uniformly at random to be {\em paired} with; according to the pairwise scheme each pair is then assigned a unique pairwise key so that they can  securely communicate with each other. We establish  critical conditions on $n, \mu$, and $K_n$ such that the resulting network has minimum node degree of at least $k$ with high probability in the limit of large network size. Our result constitutes a zero-one law for the minimum node degree of the recently introduced inhomogeneous random K-out graph model. This constitutes a crucial step towards establishing a similar zero-one law for the $k$-connectivity of the graph; i.e., for the property that the network remains connected despite the failure of any $k-1$ nodes or links. We present numerical results that indicate the usefulness of our results in selecting the parameters of the scheme in practical settings with finite number of sensors.

\end{abstract}


\section{Introduction}
\label{sec:Introduction}
\subsection{Background and Motivation}
Wireless sensor networks (WSNs) are of vital importance in numerous application domains including environmental sensing, health monitoring, surveillance and tracking \cite{Akyildiz_2002}. The affordability, scalability, low-power consumption and ease of installation has made WSNs the backbone of several emerging technologies \cite{yick2008wireless}. WSNs are often deployed in hostile environments making them susceptible to adversarial attacks and operational failures.
The limited energy, computation, and communication capabilities of WSNs precludes the use of standard cryptosystems to safeguard these networks \cite{perrig2004security}.
Eschenauer and Gligor addressed this issue of security in WSNs by introducing the notion of random key predistribution in their pioneering work \cite{Gligor_2002}. Following this work, several variants of random key predistribution emerged; e.g., see \cite{security_survey,XiaoSurvey} and the references therein. 
\par One of the widely acclaimed schemes is the random \emph{pairwise} key predistribution scheme proposed by Chan et al. in \cite{Haowen_2003}. The random pairwise key predistribution scheme comprises of two phases. First, each sensor node is paired offline with $K$ nodes chosen uniformly at random among all sensor nodes. Next, a unique pairwise key is distributed to all node pairs in which at least one of the nodes is paired to the other during the offline node-pairing step. After deployment, two sensor nodes can communicate securely if they have at least one pairwise key in common. In Section~\ref{sec:Model}, we give the precise implementation details of this scheme and its heterogeneous variant proposed in \cite{eletrebycdc2018}. The pairwise key predistribution scheme preserves the secrecy of rest of the network in case of node capture attacks, and also enables node-to-node authentication and quorum-based revocation \cite{Haowen_2003}.

\begin{figure}[!t]
\hspace{-.3cm}
\centering
\includegraphics[scale=0.14]{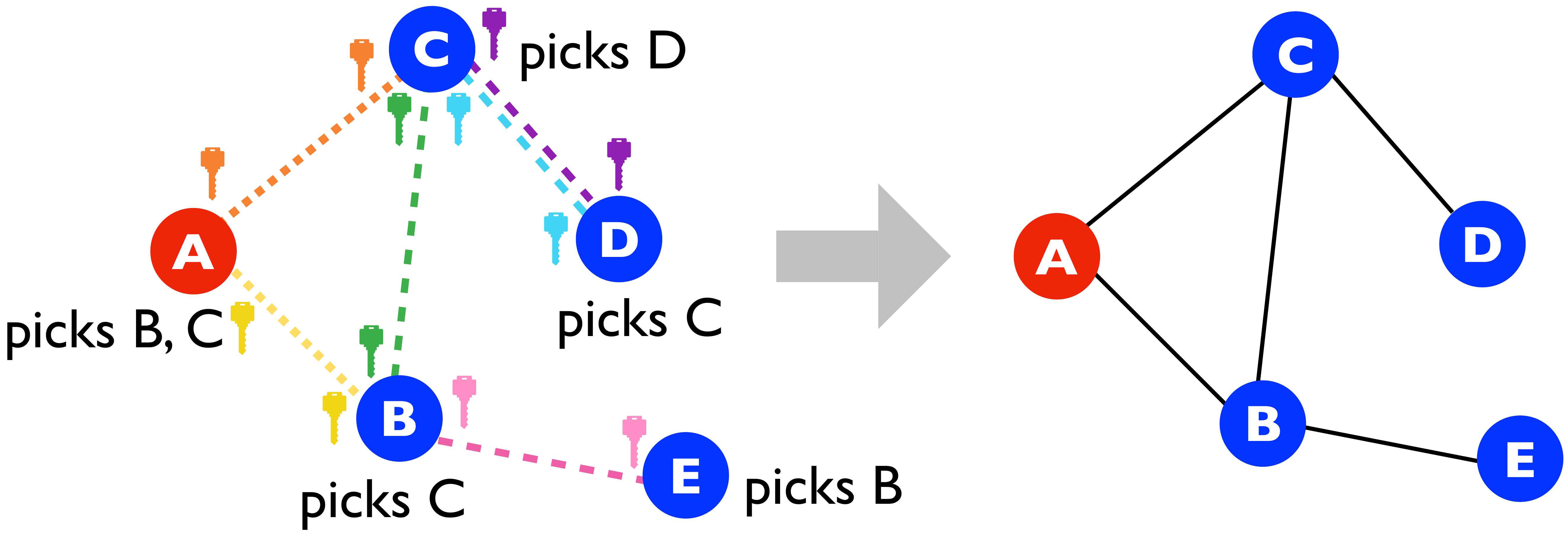} 
\vspace{4mm}
\caption{\sl 
A WSN comprising $5$ nodes secured by the heterogeneous random pairwise key predistribution scheme. 
Each type-1 (resp. type-2) node randomly picks 1 (resp. $K=2$) nodes and a unique pairwise key is given to node pairs per selection; in this example, node $A$ is type-2 and others are type-1. Two nodes can communicate if they have at least one pairwise key in common. This induces a graph with edges corresponding to node pairs which share at least one key in common.}
\vspace{-2mm}
\label{fig:pairwise}
\end{figure}
%
\par The random pairwise key predistribution scheme induces a class of random graphs known as \emph{random K-out graphs} \cite{Bollobas,FennerFrieze1982,Yagan2013Pairwise,yavuz2015toward}:
 each of the $n$ vertices is assigned $K$ arcs towards $K$ distinct vertices that are selected uniformly at random, and then the orientation of the arcs is ignored.
Let $\hhh$ denote the resulting random K-out graph. 
In \cite{Yagan2013Pairwise, FennerFrieze1982}, it was shown that if $K \geq 2$, then the resulting graph is 1-connected with high probability. More precisely, we have
\begin{equation*} 
\lim_{n \to \infty} \mathbb{P}\left[ \mathbb{H}(n;K) \text{ is connected}\right] =
\begin{cases}
1 & \mathrm{if} \quad K\geq 2, \\
0 & \mathrm{if} \quad K=1.
\end{cases}
\end{equation*} 

 Emerging real-world networks are characterized by heterogeneous nodes differing in their roles, resources, hardware limitations and connectivity requirements  \cite{Lu2008_applications, Wu2007_applications, Yarvis_2005}.
This made it necessary to develop and analyze heterogeneous variants of classical key predistribution schemes. From a theoretical stand point, this corresponds to advancing the literature pertaining to connectivity in inhomogeneous variants of classical random graph models. In order to model differing node capabilities, \cite{eletrebycdc2018} introduced a heterogeneous pairwise key predistribution scheme in which each node is classified as type-1 (respectively, type-2) with probability $\mu$ (respectively, $1-\mu)$, $0<\mu<1$. Then, each type-1 (respectively, type-2) node selects one node (respectively, $K_n$ nodes) uniformly at random from all other nodes; see Figure~\ref{fig:pairwise}. The heterogeneous pairwise key predistribution scheme induces an inhomogeneous random K-out graph, denoted $\hh$. The analysis of 1-connectivity in \cite{eletrebycdc2018} of $\hh$ yielded the rather surprising result that
\begin{equation*} 
\lim_{n \to \infty} \mathbb{P}\left[ \mathbb{H}(n;\mu,K_n) \text{ is connected}\right] =
\begin{cases}
1 & \mathrm{if}~ K_n=\omega(1), \\
<1 & \mathrm{otherwise}. 
\end{cases}
\end{equation*}  

This paper is motivated by the fact that in many applications it is desirable to have a {\em  stronger} notion of connectivity; e.g., the network remaining connected despite edge or node failures and removals. To this end, this work initiates a study on the $k$-connectivity of WSNs under the heterogeneous pairwise scheme. A network is said to be $k$-connected if it remains connected despite the removal of any $k-1$ of its nodes or edges\footnote{The notion of $k$-connectivity used in this paper refers to $k$-vertex connectivity, which is defined as the property that the graph remains connected after deletion of any $k-1$ vertices.  It is known  \cite{erdos61conn}  that a $k$-vertex connected graph is always $k$-edge connected, meaning that it will remain connected despite the removal of any $k-1$ edges. Thus, we say that a graph is $k$-connected (without explicitly referring to vertex-connectivity) to refer to the fact that it will remain connected despite the deletion of any $k-1$ vertices or edges. 
}. Moreover, in a $k$-connected graph, there are at least $k$ mutually disjoint paths between any pair of nodes. In the context of WSNs, the property of $k$-connectivity is highly desirable since it makes the network resilient to the failure of up to $k-1$ sensor nodes or $k-1$ links. Such failures could arise in practice due to operational failures, adversarial capture of nodes, or battery depletion. In addition to providing reliability against failures, $k$-connectivity facilitates the incorporation of mobile nodes with intermittent connectivity. A $k$-connected WSN can at any given time support up to $k-1$ mobile nodes without disrupting connectivity. This motivates us to study reliable connectivity, namely $k$-connectivity in networks secured by the heterogeneous pairwise key distribution scheme.

We envision that the simplicity of graph construction and unique connectivity properties position the K-out graph as a promising model for  many real-world networks in addition to  WSNs. Recently, a structure similar to the random K-out graph was proposed in \cite[Algorithm~1]{FantiDandelion2018} for generating anonymity graphs to facilitate diffusion of transaction information, thereby making the crypto-currency network robust to {\em de-anonymization} attacks. 
Given their sparse yet connected structure,   random K-out graphs can  potentially be useful also for payment channels used for scaling cryptocurrency networks such as the Lightning Network \cite{poon2016bitcoin}. 
We further motivate our work by noting that in these additional potential applications as well, it would be of interest to study the
the $k$-connectivity property and to understand the impact of heterogeneity. For example, having at least $k$ {\em mutually disjoint} paths  between every pair of nodes would be useful in  payment channels since some links could fail due to depletion of funds. 

\subsection{Main Contributions}
In this paper, we initiate the analysis of $k$-connectivity in WSNs secured by the heterogeneous pairwise key predistribution scheme which induces an inhomogeneous random K-out graph $\hh$. We establish a zero-one law for the property that the minimum node degree is at least $k$. In particular, we present conditions on  $\mu$ and $\K$ such that the resulting graph has minimum degree at least $k$ with probability approaching one (respectively, zero) constituting the one-law (respectively, zero-law), as the number of nodes gets large. 
Numerical results are presented to demonstrate the usefulness of these results in selecting the parameters of the pairwise scheme when the number of nodes is finite. 
An interesting finding is that for inhomogeneous random K-out graphs, the  number of {\em additional} edges needed to go from $1$-connectivity to $k$-connectivity with $k \geq 2$ is {\em unexpectedly} larger as compared to many other random graph models studied before; see Table \ref{table:tab1} for details.


Our result provides a crucial  step towards establishing a zero-one law for $k$-connectivity in inhomogeneous random K-out graphs. In particular, since minimum node degree being at least $k$ constitutes a necessary condition for $k$-connectivity, the zero-law established here also provides a zero-law for $k$-connectivity. In fact, taking evidence from several other random graph models \cite{erdos61conn,ZhaoYaganGligor, PenroseBook}, we conjecture that the one-law for $k$-connectivity will also be identical to the one-law established here for minimum node degree being at least $k$.  This conjecture is supported further  
through numerical results.


\subsection{Notation}  
All limits are understood with the number of nodes $n$ going to infinity. While comparing asymptotic behavior of a pair of sequences $\{a_n\},\{b_n\}$, we use $a_n = \oo(b_n)$,  $a_n = \OO(b_n)$, $a_n=\Theta(b_n)$, and $a_n = \omega(b_n)$ with their meaning in the standard Landau notation. 
 All random variables are defined on the same probability triple $(\Omega, {\cal F}, \mathbb{P})$. Probabilistic statements are made with respect to this probability measure $\mathbb{P}$, and we denote the corresponding expectation operator by $\mathbb{E}$. We let $\ii\{A\}$ denote the indicator random variable which takes the value 1 if event $A$ occurs and 0 otherwise. We say that an event holds with high probability (whp) if it holds with probability one as $n\rightarrow \infty$. We denote the cardinality of any discrete set $A$ by $|A|$ and the set of all positive integers by $\N$.  

\section{System Model}
\label{sec:Model}
\subsection{Heterogeneous random pairwise key predistribution}
The widely studied pairwise key predistribution scheme proposed by Chan et al. \cite{Haowen_2003} was extended in \cite{eletrebycdc2018} to {\em heterogeneous} settings to accommodate networks where sensors differ in their capabilities and mission requirements. 
The steps undertaken to establish secure connectivity in the heterogeneous random pairwise key predistribution scheme are as follows.
Consider a network comprising of $n$ nodes which are labeled as $i=1,2,\dots n$ and assigned unique IDs: ${\rm Id}_1, \ldots , {\rm Id}_n$. 
Each node is labeled as type-1 (respectively, type-2) with probability $\mu$ (respectively, $1-\mu$) independently from other nodes where $0<\mu<1$. In the (offline) \emph{initialization} phase, each type-1 (respectively, type-2) node selects $K_1$ (respectively, $K_2$) other nodes chosen uniformly at random. 
The homogeneous pairwise scheme \cite{Haowen_2003} corresponds to the case where $\mu=0$ and all nodes choose exactly $K$ nodes to be paired with.


Define $\nodes:=\{1,2,\dots,n\}$ and $\nodes_{-i}:=\{1,2, \dots, n\}\setminus i$.
For each $i \in \nodes$, let $\G_{n,i} \subseteq \nodes_{-i}$ denote the subset of nodes  selected by node $i$ from $\nodes_{-i}$ uniformly at random. With
 $t_i \in \{1,2\}$ denoting the type of node~$i$, 
we have for any  $A \subseteq {\cal N}_{-i}$
\[
\pr[{ \G_{n,i} = A~|~t_i=\ell }] = \left \{
\begin{array}{ll}
{{n-1}\choose{K_\ell}}^{-1} & \mbox{if $|A|=K_\ell$}, \\
              &                   \\
0             & \mbox{otherwise.} \\
\end{array}
\right .
\]
We further assume that $\Gamma_{n,1}, \ldots , \Gamma_{n,n}$ are mutually independent given the types of nodes. 

Once the offline pairing process has been  completed, 
we insert key rings 
$ \Sigma_{n,1},\dots,\Sigma_{n,n}$ in the memory modules as follows. If
 $   i \in \G_{n,j} \lor j \in \G_{n,i}$, i.e., either node $i$ selects node $j$ or node $j$ selects node $i$  or both,
 we generate a pairwise  key $\omega_{ij}$ and store this key and the corresponding node IDs in the memory modules of both nodes $i$ and $j$. It is important to note that the key $\omega_{ij}$ is  assigned \emph{exclusively} to nodes $i$ and $j$ to be used in securing the communication between them. This strategy of assigning unique keys to nodes which were paired during the offline node-pairing process is the reason why this approach is called the \emph{pairwise} key predistribution scheme. Having unique, pairwise keys brings several advantages including distributed node-to-node authentication  and resilience to node capture attacks    \cite{Haowen_2003}. 

In the post-deployment \emph{key-setup} phase, nodes first broadcast their IDs to their neighbors following which each node searches for the corresponding IDs in their key rings. Finally, node pairs wishing to communicate verify each others' identities through a cryptographic handshake \cite{Haowen_2003}. Thus, by construction, the pairwise key predistribution facilitates node-to-node authentication.

In the rest of this paper, we assume $K_1=1$ and $K_2 \geq 2$ as in \cite{eletrebycdc2018} for  simplicity. The more general cases with arbitrary number of node types and arbitrary scheme parameters $K_1, K_2, \ldots$, should be studied in a separate paper. We assume that $0<\mu<1$ is fixed and $K_2$ scales with $n$. From here onward, let $K_n$ denote the scaling of $K_2$ with $n$. 
In
 \cite{eletrebycdc2018}, the $1$-connectivity of the network under this setting was studied. In particular, it was shown that the network is connected whp only if $K_n$ grows unboundedly large as $n \to \infty$ (irrespective of $\mu$). 
 This was in stark contrast with the results for the homogeneous case where it is known that the network is connected whp if $K \geq 2$ (with $K$ being the number of choices made by every node). 
 The main goal of this paper is to initiate a study on the $k$-connectivity of the network under the same setting; e.g., by revealing conditions on $K_n$ and $\mu$ required for the network to be $k$-connected whp.

\subsection{Inhomogeneous Random K-out graph}
A WSN comprising of $n$ sensors secured by the heterogeneous pairwise key predistribution scheme can be modeled by an inhomogeneous random K-out graph defined as follows.
Two distinct nodes $i$
and $j$ are said to be adjacent, written $i \sim j$, if and only if they have at least one common key in their respective key rings as defined above. 
Equivalently, nodes $i$
and $j$ are adjacent if either node picks the other or both; i.e.,
\begin{align}
\vspace{-2mm}
i \sim j ~~\quad \mbox{if} ~~~\quad j \in \G_{n,i} ~
\vee~ i \in \G_{n,j}. 
\label{eq:Adjacency}
\end{align}
Thus, given a set of $n$ nodes,  the adjacency condition (\ref{eq:Adjacency})  gives a precise edge construction on the vertex set $\{1,2,\dots,n\}$. We denote the graph constructed using  (\ref{eq:Adjacency}) as $\hh$. Further, let $\kk$ denote the average number of selections per node given by $\mu+(1-\mu)K_n$. 
We note that if $\mu=0$,  the inhomogeneous random K-out  graph becomes equivalent to the homogeneous random K-out  graph \cite{Yagan2013Pairwise, FennerFrieze1982, yavuz2015toward}.

\section{Results and Discussion}
\label{sec:mainresult}

In this section, we present our main technical result: a zero-one law for the minimum node degree of the inhomogeneous random K-out graph 
being at least $k$. We then provide a discussion on the implications of our main result, and we explain why it is expected to pave the way to establish a similar zero-one law for $k$-connectivity. Finally, we provide experimental results that demonstrate the usefulness of our result in the finite node regime and support our conjecture that an analog of Theorem \ref{theorem:th1} holds  for $k$-connectivity.

\subsection{Main results}
We refer to any mapping $K:\N \rightarrow \N$ as a {\em scaling} if it satisfies the condition 
\[
2 \leq K_n < n, \quad  n=2, 3, \ldots .
\] 
Our main result is presented next.
\begin{theorem}
{\sl 
 Consider a scaling $K:\N \rightarrow \N$ and $\mu$ such that $0<\mu<1$. 
With $\kk = \mu+(1-\mu)K_n$ and a positive integer $k\geq2$ let the sequence $\g: \N\rightarrow \R$ be  defined through
\begin{align}
\kk=\log n +(k-2)\log \log n+\g_n, 
\label{eq:hyp}
\end{align}
for all $n=2, 3, \ldots$. Then, we have
\begin{align*}
   \limit \pr\left[\begin{array}{ll}
   \text{Min.~node degree of }    &  \\
     \text{$\hh$ is  $~\geq k$} & 
  \end{array}
  \hspace{-4mm}\right]
  = \begin{cases}
                                   1 & \textrm{if } \limit \g_n =+\infty, \\
                                   & \\
                                   0 & \textrm{if } \limit \g_n =-\infty. \\
  \end{cases}
\end{align*}
\label{theorem:th1}
}
\end{theorem}

Theorem~\ref{theorem:th1} establishes scaling conditions on $\kk$ (i.e., on $K_n$ and $\mu$) such that the minimum node degree of the inhomogeneous random K-out graph $\hh$ is at least $k$, {\em and} less than $k$, respectively, with high probability
as the number of nodes approaches infinity.
Put differently, it establishes a {\em zero-one law} for the property that $\hh$ has minimum degree of at least $k$. 
An immediate consequence of Theorem~\ref{theorem:th1} is that $\kk$ must scale as $\Omega(\log n)$ for the minimum node degree to be at least $k$.
The scaling condition (\ref{eq:hyp}) in Theorem~\ref{theorem:th1} can be equivalently expressed in terms of the network parameters $\mu$ and $K_n$ as follows
\begin{align}
K_n=\dfrac{\log n +(k-2)\log \log n}{1-\mu}+\g_n. \label{eq:hyp2}
\end{align}

The proof of Theorem~\ref{theorem:th1} is based  on  applying the method of first and second moments \cite{JansonLuczakRucinski} to a random variable  counting the number of nodes with degree less than $k$. The dichotomous results with $\g_n$ approaching $+\infty$ and $-\infty$ can be intuitively viewed as a consequence of the expected number of nodes with degree less than $k$ approaching $\infty$ (respectively, 0) as $\g_n$ goes to $\infty$ (respectively, $-\infty$). However, the complex inter-dependencies in the node degrees make this analysis quite challenging.
In particular, the second moment analysis requires careful consideration of several cases for the node degree of a pair of nodes and the selections made by them. Due to space constraints, we provide a brief sketch of the proof for Theorem~\ref{theorem:th1} in Section~\ref{sec:proofTh1}; all details can be found in the Appendix.

\subsection{Discussion}
Recall that a $k$-connected graph remains connected upon deletion of any $k-1$ nodes or edges. Consequently, for a graph to be $k$-connected, the minimum node degree for the graph must be at least $k$. Thus, Theorem~\ref{theorem:th1} also provides a necessary condition for $k$-connectivity in networks secured by the heterogeneous pairwise key predistribution scheme. In particular, under scaling (\ref{eq:hyp}), Theorem~\ref{theorem:th1} automatically gives the corresponding zero-law for $k$-connectivity. Furthermore, in most random graph models including Erd\H{o}s-R\'enyi graphs \cite{erdos61conn}, random key graphs \cite{ZhaoYaganGligor} and random geometric graphs \cite{PenroseBook}, the conditions required to ensure a minimum node degree of at least $k$ are shown to be sufficient to make the graph $k$-connected whp. This is done 
by showing that it is highly unlikely for the corresponding graph to be \emph{not} $k$-connected if all nodes have at least $k$ neighbors. In light of this (as well as our numerical studies presented in Section~\ref{subsec:numerical}), we conjecture the following zero-one law for $k$-connectivity. 
\begin{conjecture}
{\sl \footnote{This conjecture has been proved: {M. Sood}, O. Ya\u{g}an, \qq{\emph{Zero-one law for $k$-connectivity  in Inhomogeneous Random $K$-out Graphs}}, 2019.}
Consider a scaling $K:\N \rightarrow \N$ and $\mu$ such that $0<\mu<1$. 
With 
a positive integer $k\geq2$ let the sequence $\g: \N\rightarrow \R$ be  defined through
(\ref{eq:hyp}). Then, we have
\begin{align*}
   \limit \pr\left[\begin{array}{ll}
   \textrm{$\hh$ is }    &  \\
     \textrm{$k$-connected} & 
  \end{array}
  \hspace{-4mm}\right]
  = \begin{cases}
                                   1 & \textrm{if } \limit \g_n =+\infty, \\
                                   & \\
                                   0 & \textrm{if } \limit \g_n =-\infty. \\
  \end{cases}
\end{align*}
\label{conjecture:c1}
}
\end{conjecture}

We now discuss the  implications of Theorem \ref{theorem:th1} on connectivity for heterogeneous pairwise predistribution scheme and compare the results with other key predistribution schemes.  Table~\ref{table:tab1}
summarizes the mean node degree requirements for having $1$-connectivity and $k$-connectivity for homogeneous and inhomogeneous versions of random K-out graphs \cite{eletrebycdc2018, FennerFrieze1982} and {\em random key graphs} induced by the EG scheme \cite{Gligor_2002,yagan2012zero,Yagan/Inhomogeneous,8606999}. For the inhomogeneous versions, the table entries correspond to the mean degree of the {\em least connected} node type.
We also added the corresponding results for Erd\H{o}s-R\'enyi graphs \cite{erdos61conn} for comparison. An interesting observation is that for
the inhomogeneous random K-out graph, going from $1$-connectivity to $k$-connectivity requires an increase of $\log n +(k-2)\log \log n$ in the mean degree. This is much larger than what is required (i.e., $(k-1) \log \log n$) in the other models seen in Table \ref{table:tab1}.
In addition, we see that the homogeneous pairwise key predistribution scheme incurs the least overhead in terms of the edges and keys required to achieve 1-connectivity or $k$-connectivity. For instance, if all nodes select two neighbors, the resulting network is securely 2-connected. However, the analysis of $1$-connectivity in \cite{eletrebycdc2018} and Theorem \ref{theorem:th1} together suggest that the perceived benefits of the pairwise scheme reduce in the face of heterogeneity when $K_1=1$, i.e., when a positive fraction of nodes picks just one other node to be paired with. 


\begin{table}[htbp]
\vspace{2mm}
\begin{center}
 \begin{tabular}{|L| M| N|} 
 \hline
\textbf{Random graph} & \textbf{$\pmb{1}$-connectivity} &   \textbf{$\pmb{k}$-connectivity}, $k\geq2$ \\
 \hline\hline
  Homogeneous K-out &  $4$ & $2k$ \\ 
 \hline
  Inhomogeneous K-out &  $\omega(1)$ & $\log n + (k-2) \log \log n + \omega(1)$  \\ 
 \hline
  Homogeneous random key &  $\log n + \omega(1)$ & $\log n + (k-1) \log \log n + \omega(1)$ \\
 \hline
  Inhomogeneous random key &  $\log n + \omega(1)$ & $\log n + (k-1) \log \log n + \omega(1)$ \\
 \hline
  Erd\H{o}s-R\'enyi &  $\log n + \omega(1)$ & $\log n + (k-1) \log \log n + \omega(1)$  \\
 \hline
\end{tabular}
\end{center}
\caption{\sl Comparing the mean node degree necessary for 1-connectivity and $k$-connectivity in several random graph models. For inhomogeneous K-out and inhomogeneous random key graphs, the values given in the table correspond to the mean degree for the {\em least} connected node type.}
\label{table:tab1}
\vspace{-3mm}
\end{table}





\subsection{Simulation results}
This section presents empirical studies which probe the applicability of Theorem~\ref{theorem:th1} in the non-asymptotic regime
with finite number of sensors. For each experiment, we fix the number of nodes at $n=500$ and generate $1000$ independent realizations of the inhomogeneous random K-out graph $\hh$. 
To obtain the empirical probability of minimum node degree being at least $k$, we divide the number of instances for which the generated graph has minimum node degree of at least $k$ by the total number $1000$ of instances generated. Likewise, we compute the empirical probability for $k$-connectivity by counting the number of times (out of $1000$) for which the resulting graph is $k$-connected.

\begin{figure}[t]
\vspace{1mm}
     \centering
     \begin{subfigure}[b]{0.5\textwidth}
         \centering
         \includegraphics[scale=0.34]{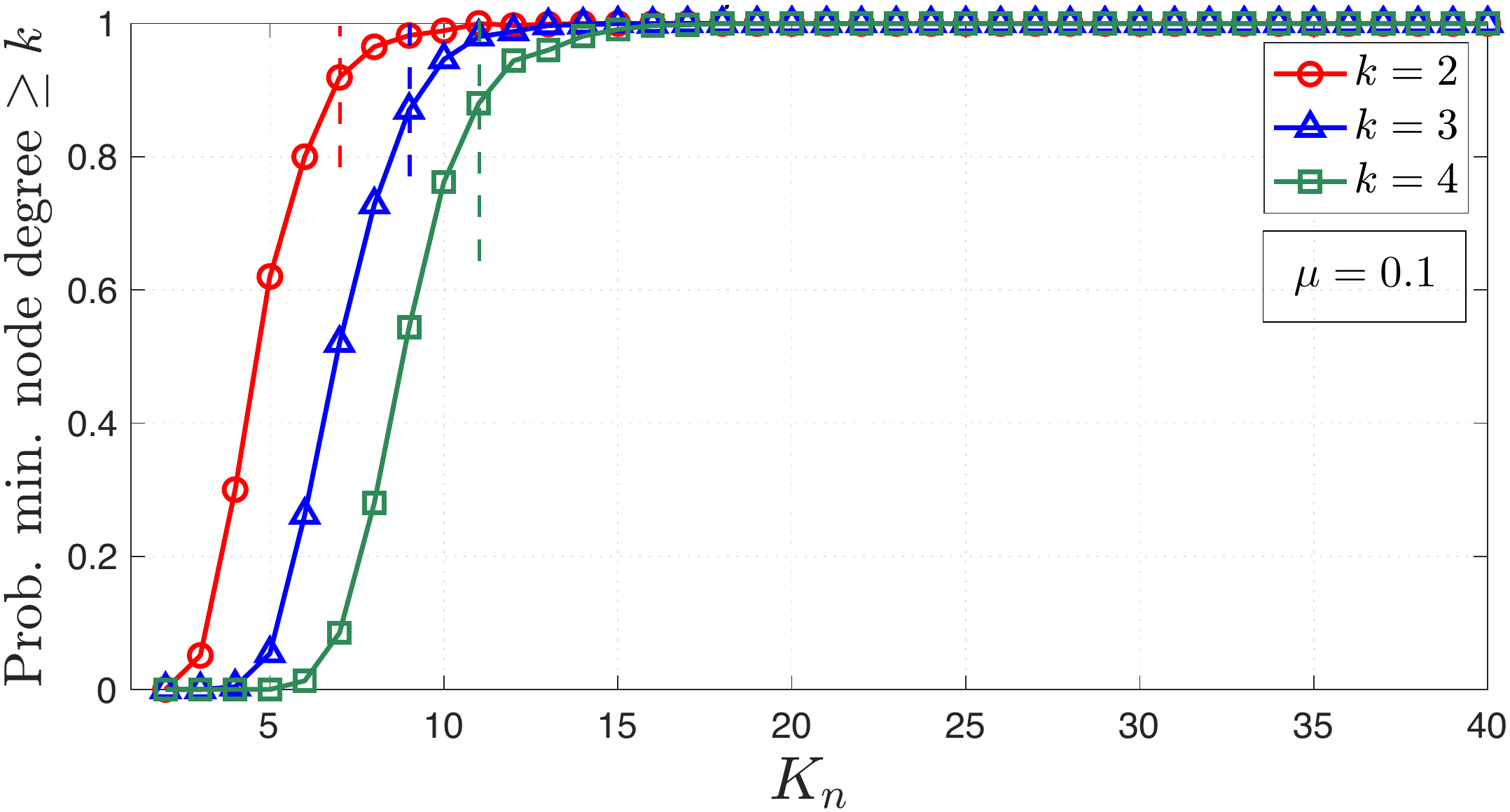}
         \label{fig:mu1}
     \end{subfigure}
     \hfill
     \vspace{.5mm}
     \begin{subfigure}[b]{0.5\textwidth}
         \centering
         \includegraphics[scale=0.34]{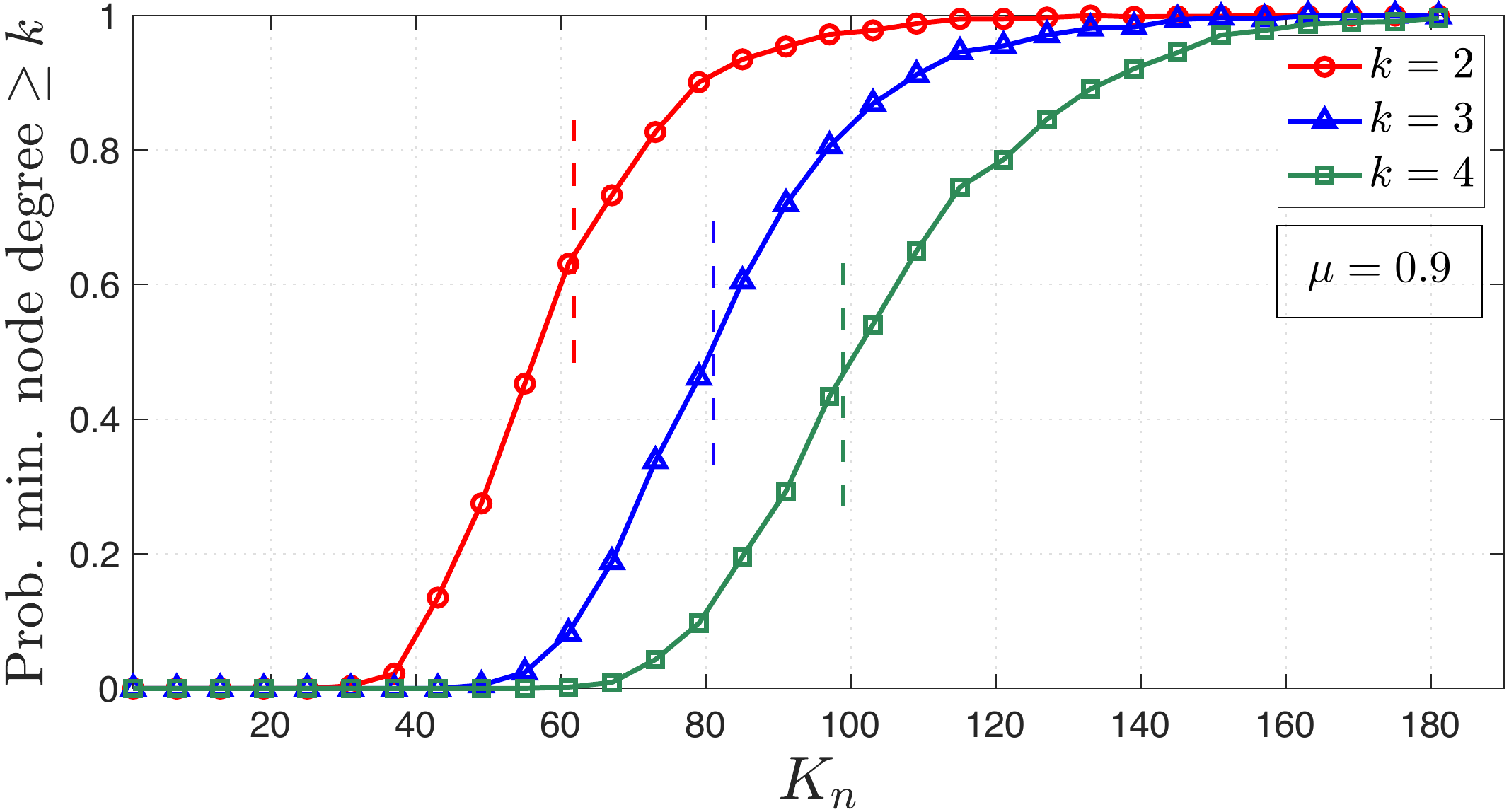} 
         \label{fig:mu9}
     \end{subfigure}
        \caption{\sl Empirical probability (computed by averaging $1000$ independent experiments for each data point) for the minimum node degree being no less than $k$ as a function of $K_{n}$ for $n=500$ and $\mu=0.1, 0.9$.  \vspace{-3mm}}
        \label{fig:mu1_9}
\end{figure}

Through simulations, we study the impact of  the probability of assignment to type-1 ($\mu$) and the number of selections made by type-2 nodes ($K_n$) on the probability that minimum node degree is at least $k$, for various $k$ values. A smaller value of $\mu$ corresponds to a network dominated by type-2 nodes and thus the resulting random graph $\hh$ has a larger probability of having a minimum node degree of at least $k$. 
Figure~\ref{fig:mu1_9} illustrates this phenomenon by comparing the cases with $\mu=0.1 $  and $\mu=0.9$. Moreover, for $k_1\geq k_2 \geq2$, the event that a graph has a minimum node degree of at least $k_1$ is a subset of the event that a graph has a minimum node degree greater than or equal to $k_2$. Thus, 
we see an upwards shift in the curves when going from $k=4$ to $k=2$.

In each figure, we also plot a vertical-dashed line corresponding to the critical threshold of having minimum node degree at least $k$ asserted by Theorem~\ref{theorem:th1} through scaling condition (\ref{eq:hyp2}). Namely, vertical dashed lines represent the $K_n$ value satisfying
\begin{align}
K_n= \left \lceil \frac{\log n +(k-2)\log \log n}{1-\mu}
\right \rceil \label{eq:thresh}.
\end{align}
The desired probability of minimum node degree being no less than $k$ is 0 for small values of $K_n$ and increases sharply to 1 in a neighborhood of the threshold described by (\ref{eq:thresh}). Thus, we find a good agreement between our numerical results and Theorem~\ref{theorem:th1} even when the number of nodes is finite.

\begin{figure}[h]
     \centering
     \begin{subfigure}[b]{0.5\textwidth}
         \centering
         \includegraphics[scale=0.34]{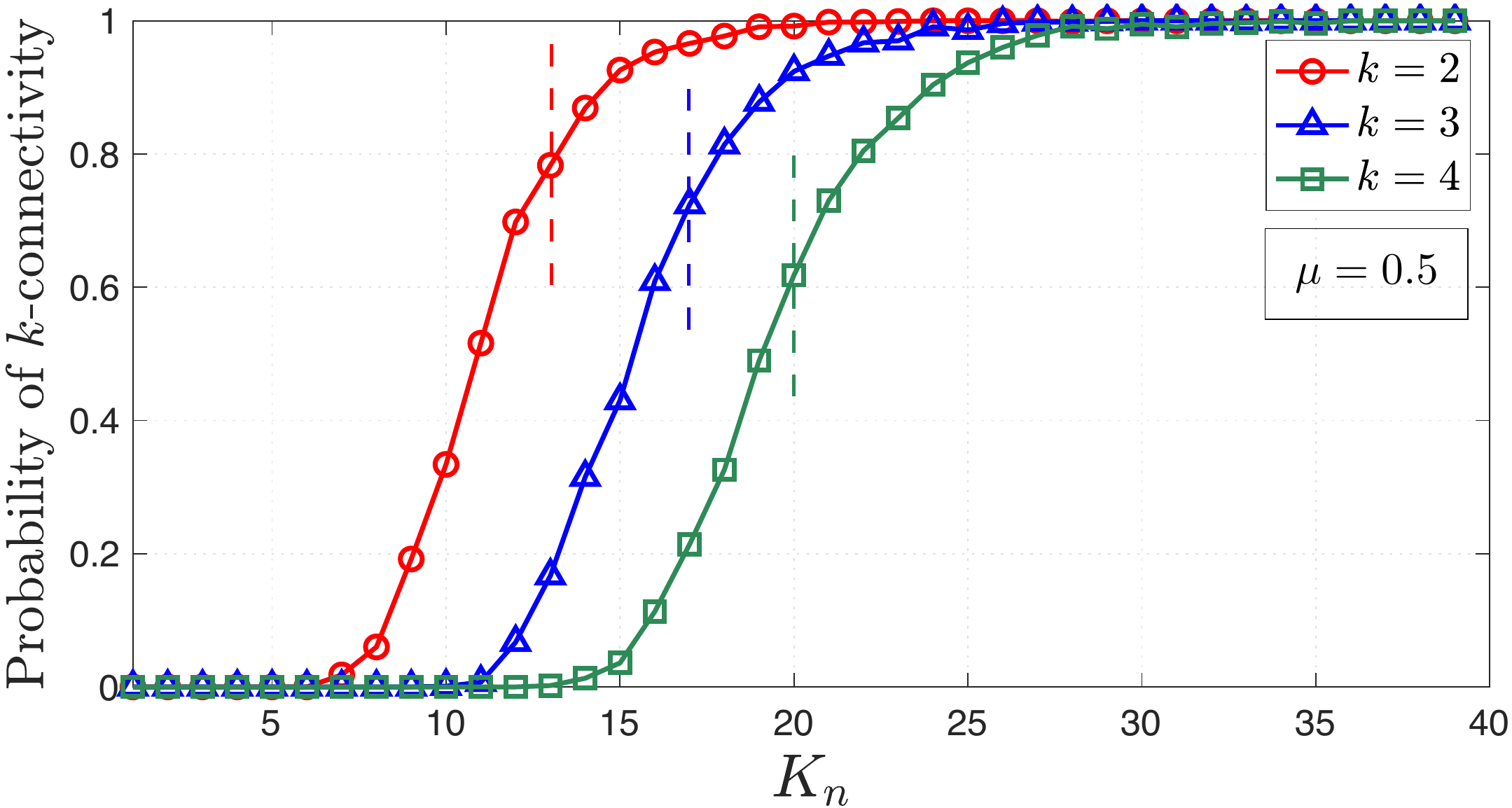}
         \label{fig:mu5mnd}
     \end{subfigure}
     \hfill
          \vspace{.5mm}
     \begin{subfigure}[b]{0.5\textwidth}
         \centering
         \includegraphics[scale=0.34]{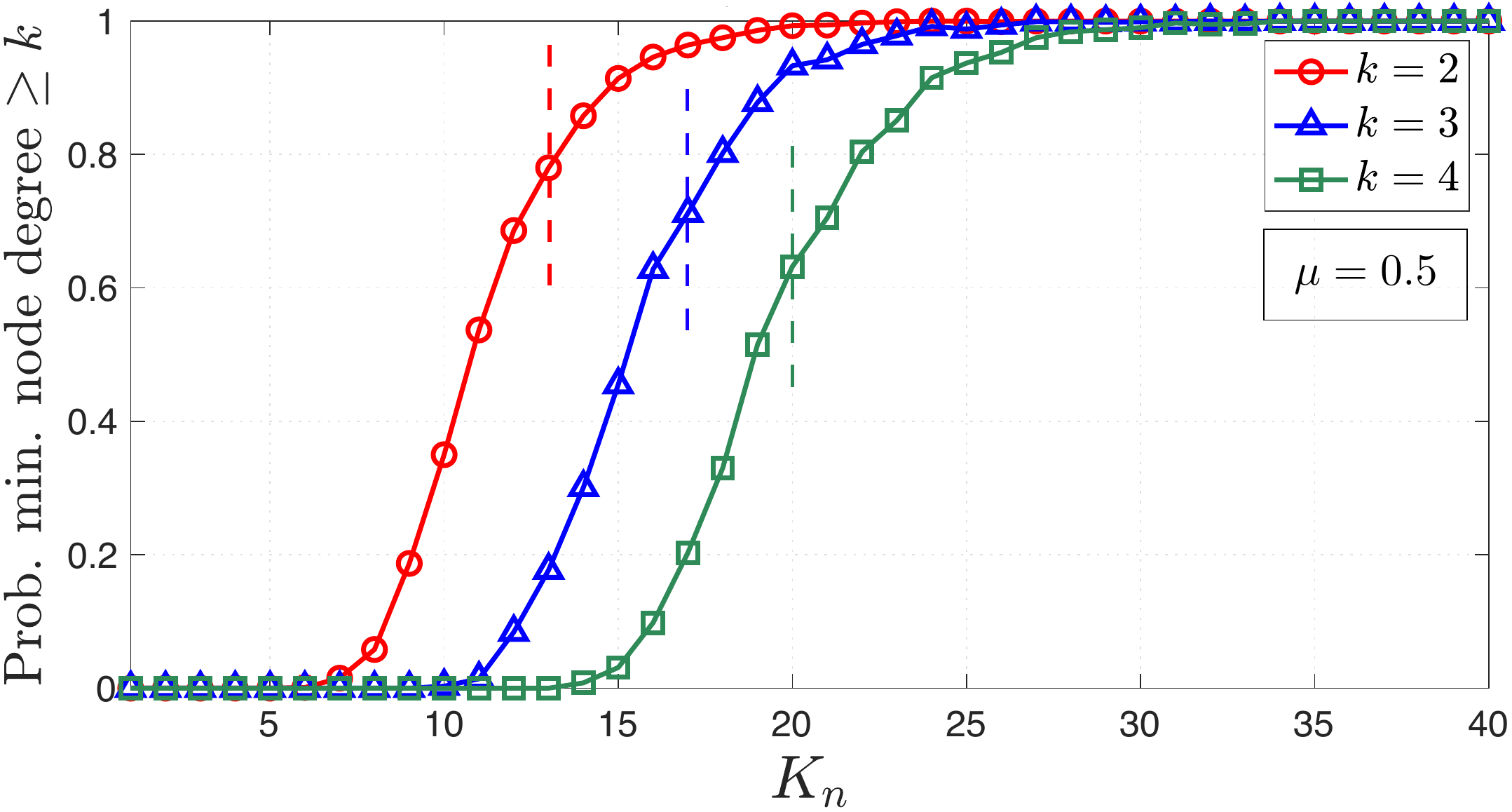} 
         \label{fig:mu5kconn}
     \end{subfigure}
        \caption{\sl Empirical probability 
        (computed by averaging $1000$ independent experiments for each data point) of $k$-connectivity and minimum node degree being no less than $k$ as a function of $K_{n}$ for $n=500$, $\mu=0.5$ and $k=2,3,4$.  \vspace{-3mm}}
        \label{fig:mu5}
\end{figure}

Finally, we  test the validity of  Conjecture~\ref{conjecture:c1} for $k$-connectivity. 
Figure~\ref{fig:mu5} presents the empirical probability of $k$-connectivity and minimum node degree being no less than $k$ as $k$ varies from $2$ to $4$ and $\mu=0.5$. We observe a striking similarity between the plots for minimum node degree and $k$-connectivity. Although not shown here for brevity, the same phenomenon is observed in all experiments with different parameter settings for $\mu$ and $k$. This provides strong evidence in favour of  minimum node degree being at least $k$ and $k$-connectivity being {\em asymptotically} equivalent properties in inhomogeneous random K-out graphs.

\label{subsec:numerical}
\section{Sketch of the Proof of Theorem~\ref{theorem:th1}}
\label{sec:proofTh1}
In this section, we provide a brief outline for the proof of Theorem~\ref{theorem:th1}; all details 
can be found in the Appendix.
Our proof  employs the method of first and second moments \cite{JansonLuczakRucinski} applied to count variables representing the number of nodes in $\hh$ with degree less than $k$.  
With any $d=0, 1, \ldots$, let $Z_{n,d}$ denote the number of nodes in $\hh$ with degree $d$. 
In other words, we let $Z_{n,d}=\sum_{i=1}^{n}\ii\{\deg(v_i) = d\}$,
where $\deg(v_i)$ is the degree of node $v_i$.
To provide an intuitive explanation for the dichotomy arising in Theorem~\ref{theorem:th1} (depending on the limiting value of the sequence $\g_n$), we present the next result on the mean value of  $Z_{n,d}$
under an additional technical condition on $\g_n$; as can be seen in the detailed proof of Theorem~\ref{theorem:th1}  given in the Appendix, this condition is not needed for our  result to hold. 


\begin{lemma}
{\sl With $\g_n$ defined through (\ref{eq:hyp}), let $|\g_n|=\oo{(\log n)}$. Then, the expected number of nodes with degree $d$ satisfies
\begin{align*}
\mathbb{E}\left[Z_{n,d}\right] = \Theta(1) \exp\left\{-(k-1-d)\log \log n - \gamma_n\right\}.
\end{align*}
\vspace{-3mm}
\label{lemma:simplified}
}
\end{lemma}

In order to see how the zero-one law arises in Theorem~\ref{theorem:th1}, substitute $d=k-1$ in Lemma~\ref{lemma:simplified}. We see that $\E[Z_{n,k-1}] = \Theta(e^{-\gamma_n})$ so that 
it approaches $0$ (respectively, $\infty$) as $\g_n$ tends to $\infty$ (respectively, $-\infty$). This dichotomy forms the basis of defining $\gamma_n$ in the {\em critical} scaling condition (\ref{eq:hyp}), and the accompanying zero-one law given in Theorem \ref{theorem:th1}.

To establish the one-law, we use the Markov inequality applied to the integer-valued random variable $Z_{n,d}$ leading to
$\pr[Z_{n,d} \leq 1] \leq \E[Z_{n,d}]$. 
This gives 
\begin{align}
     \pr[Z_{n,d} = 0] &= 1 - \pr[Z_{n,d} \leq 1] \geq 1-\E[Z_{n,d}]. \label{eq:onelaw}
\end{align}
From Lemma \ref{lemma:simplified},
when $ \g_n \to \infty$, we see that $\E[Z_{n,d}] \to 0$ 
for all $d =1,2,\dots,k-1$. Using this in  
(\ref{eq:onelaw}), we get
\[
\limit \pr[Z_{n,d} = 0] = 1, \quad d=0, 1, \ldots k-1
\]
which is equivalent to the one-law
\[
\limit  \pr\left[ 
   \textrm{Min.~node degree of }    
     \textrm{$\hh$ is  $\geq k$}  
\right] = 1.
  \]


In order to obtain a zero-law, we invoke the method of second moments. Using the Cauchy-Schwarz inequality it can be shown \cite{JansonLuczakRucinski} that 
\begin{align}
    \pr[Z_{n,d} \neq 0]\geq \frac{(\E[Z_{n,d}])^2}{\E[Z_{n,d}^2]}. \label{eq:zerolawconditionprob}
\end{align}
From exchangeability of the indicator random variables $\ii\{\deg(v_i)=d\}$, $i=1,2,\dots,n$, we have
\begin{align}
\hspace{-2mm}\frac{\E[Z_{n,d}^2]}{(\E[Z_{n,d}])^2}\hspace{-.5mm}=\hspace{-1mm} \frac{1}{\E[Z_{n,d}]} \hspace{-.5mm}+\hspace{-.5mm}\frac{n-1}{n}
\dfrac{\pr[\deg(v_1)\hspace{-.5mm}=\deg(v_2)=d]}{{\left(\pr[\deg(v_1)=d]\right)^2}}.
\label{eq:zerolawcondition}
\end{align}

In view of (\ref{eq:zerolawconditionprob})-(\ref{eq:zerolawcondition}), we get $\pr[Z_{n,d} \neq 0] \to 1$ if
the following two results are established.
\begin{enumerate}
    \item $\limit \E[Z_{n,d}] =\infty$,
    \item ${\underset{n \rightarrow \infty}\limsup}\dfrac{\pr[\deg(v_1)=\deg(v_2)=d]}{{\left(\pr[\deg(v_1)=d]\right)^2}} \leq 1$.
\end{enumerate}
\vspace{2mm}

The zero-law in Theorem \ref{theorem:th1} follows upon establishing that both conditions hold with $d=k-1$ when $ \g_n \to -\infty$.  
The first condition  follows from Lemma~\ref{lemma:simplified} when $ \g_n \to -\infty$ and $d=k-1$. The proof of the second condition is involved due to the
degrees of $v_1$, $v_2$ being correlated in several ways. First, if $v_1$ picks $v_2$ or vice versa, the degrees of both nodes are affected. Second, if one of the remaining $n-2$ nodes is known to pick $v_1$, the chances of  $v_2$ being picked by the same node decreases; the exact correlations will also depend on the {\em type} of that third node. These complex correlations among node degrees makes it necessary to consider several realizations of the graph which result in a particular degree for $v_1,v_2$.  We direct readers to the Appendix for a complete proof of this result and other details of Theorem \ref{theorem:th1}. 

\section{Conclusion}
\label{sec:conc}

In this work we initiate the analysis of  $k$-connectivity in WSNs secured by the heterogeneous pairwise key predistribution scheme; 
 $k$-connectivity makes the network resilient to failure of up to $k-1$ nodes or links. 
Our main result provides {\em critical} conditions on the network and scheme parameters such that with high probability
 every sensor has at least $k$ other sensors that it can securely communicate with (when the  number of nodes gets large). Through numerical simulations, this result is shown to be useful in selecting scheme parameters  in the finite node regime. An interesting finding is that for inhomogeneous random K-out graphs, which are induced under the heterogeneous pairwise scheme, the number of {\em additional} edges needed to go from $1$-connectivity to $k$-connectivity with $k \geq 2$ is much larger than that seen in most other random graph models studied before; see Table \ref{table:tab1} for details.

This paper completes a necessary crucial step towards establishing the $k$-connectivity under the heterogeneous pairwise scheme. In fact, our numerical results suggest that an analog of our main result with the same critical scaling applies also for $k$-connectivity of the network. 
An immediate direction for future work is to prove this conjecture on $k$-connectivity. It would also be interesting to extend our analysis to a more generalized heterogeneous pairwise key predistribution scheme with $r>2$ node types and arbitrary scheme parameters $K_1, K_2, \ldots, K_r$ associated with each node type.


\bibliographystyle{IEEEtran}

\bibliography{IEEEabrv,references}
\onecolumn


\begin{center}
    {\large \bf  Appendix}\\
\end{center}

In this Section, we present a detailed proof of Theorem~\ref{theorem:th1}. We remind the readers that $\hh$ is the inhomogeneous random K-out graph induced by the heterogeneous pairwise key predistribution scheme. Recall that for each $i \in \nodes$, $\G_{n,i} \subseteq \nodes_{-i}$ denotes the subset of nodes selected by node $i$. In order to make the dependence of $\G_{n,i}$ on the model parameters more explicit we instead use the notation $\G_{n,i}(\mu, K_{n})$. In this section, we first present some preliminary results and provide a road-map for the proof of Theorem~\ref{theorem:th1}.
\setcounter{equation}{0}
\setcounter{section}{0}
\renewcommand{\thesection}{\Alph{section}}
\renewcommand \thesubsection{\Alph{section}.\Roman{subsection}}
\renewcommand{\theequation}{\thesection.\arabic{equation}}
\renewcommand{\thetheorem}{\thesection.\arabic{theorem}}
\section{Preliminaries}
\label{sec:apreliminaries}
\subsection{Mean node degree in $\hh$}
Let $\kk$ denote the mean number of edges that each node chooses to draw. Conditioning on the class of node~$i$, we get
\begin{align}
\kk&=\mu+(1-\mu)K_n. 
\end{align}
The probability that node~$i$ picks node~$j$  where $i,j\in \nodes$ depends on the type of node~$i$ and is given by
\begin{align}
\pr[j \in \G_{n,i}(\mu, K_{n})]=\mu\dfrac{1}{n-1}+(1-\mu)\dfrac{K_n}{n-1}=\dfrac{\kk}{n-1}.
\end{align}
Recall that each node draws edges to other nodes independently of other nodes. Let $i \sim j$ denote the event that node~$i$ can securely communicate with node~$j$. For $i \sim j$ to occur, either node $i$ selects node $j$ or node $i$ selects node $j$ or both select each other. This gives
\begin{align}
\pr[i \sim j]&= 1 - (1-\pr[i \in \G_{n,j}(\mu, K_{n})])(1-\pr[j \in \G_{n,i}(\mu, K_{n})]),\nonumber \\
&=1-\left(1-\dfrac{\kk}{n-1}\right)^2,\nonumber\\
&=\dfrac{2\kk}{n-1}-\left(\dfrac{\kk}{n-1}\right)^2.
\end{align}
Consequently, the mean degree of node~$i$ can be computed as follows.
\begin{align}
\E\left[\sum_{j\in \nodes_{-i}}\nonumber \ii\{i \sim j\}\right] &= (n-1)\pr[i \sim j],\\
&=2\kk-\dfrac{\kk^2}{n-1}. \label{eq:avgnd}
\end{align}
An immediate consequence of (\ref{eq:avgnd}) is that if the mean number of edges drawn by a node $\kk$ scales as $\oo(n)$, then the resulting mean node degree is $2\kk(1-\oo(1))$.
\subsection{Road-map for proof of Theorem~\ref{theorem:th1}}
The proof of Theorem~\ref{theorem:th1} hinges on the method of moments \cite{JansonLuczakRucinski} applied to count variables representing the number of nodes in $\hh$ with degree less than $k$. We separately enumerate the steps leading to the one-law and zero-law in Theorem~\ref{theorem:th1}.
\subsubsection{Establishing the one-law in Theorem~\ref{theorem:th1}}
 Let $\xd$ denote the number of nodes in $\hh$ with degree $d$ where $d=1,\ldots,k-1$.
In other words, we let $\xd=\sum_{i=1}^{n}\ii\{\deg(v_i) = d\}$,
where $\deg(v_i)$ is the degree of node $v_i$.  Since each node makes at least one selection, no node can have degree zero.
To establish the one-law, we use the method of first moment in which Markov inequality is applied to the integer-valued random variable $\xd$ yielding
\begin{align}
  \pr[\xd \leq 1] \leq \E\left[\xd\right].  
\end{align}
This gives 
\begin{align}
     \pr\left[\xd = 0\right] &= 1 - \pr\left[\xd \leq 1\right],\nonumber\\
     &\geq 1-\E\left[\xd\right]. \label{eq:markovineq}
\end{align}
Thus, if we establish that $\E\left[\xd\right] \to 0$ for all $d =1,2,\dots,k-1$ when $ \g_n \to \infty$, then we can obtain  
\begin{align}
\limit \pr[\xd = 0] = 1, \quad d=0, 1, \ldots, k-1
 \label{eq:ac1}.\end{align}
This in turn gives
\[
\limit  \pr\left[ 
   \textrm{Min.~node degree of }    
     \textrm{$\hh$ is  $\geq k$}  
\right] = 1.
  \]
  Therefore, in order to establish the one-law in Theorem~\ref{theorem:th1}, it suffices to show that the following proposition holds.
  \begin{proposition}[Establishing one-law in Theorem~\ref{theorem:th1}]
  {\sl 
  Consider a scaling $K_n:\N \rightarrow \N$, $0<\mu<1$ with $\kk = \mu+(1-\mu)K_n$, a positive integer $k\geq2$ and sequence $\g_n$ defined through
    \begin{align}
    \kk=\log n +(k-2)\log \log n+\g_n, \text{  for all }  n=2, 3, \ldots.
    \nonumber
    \end{align}
    \label{prop:aone-law}
If $\g_n \rightarrow \infty$, then the expected number of nodes in $\hh$ with degree $d$ where $d=1,2,\dots,k-1$ approaches $0$, i.e., 
  \begin{align}
     \limit \E \left[\xd \right] = 0 \text{\qquad if \qquad} \g_n \rightarrow \infty. \nonumber
  \end{align}
  }
  \end{proposition}

  
  \subsubsection{Establishing the zero-law in Theorem~\ref{theorem:th1}}
We saw above how the one-law can be proved as a consequence of the Markov inequality applied to an integer-valued random variable counting the number of nodes with degree less than $k$. This technique is referred to as the method of first moment. Next, we describe an outline for the proof of zero-law using the method of second moment. We denote the number of type-1 nodes with degree $k-1$ by $\zl$; i.e., 
\begin{align}
\zl&:=\sum_{l=1}^{n}\nonumber \ii\{\deg(v_i)=k-1, t_i=1\},
\end{align}
Following the method of second moment,
we obtain a lower bound on $\pr[\zl \neq 0]$ as follows. 
\begin{align}
    (\E [\zl])^2&= \left(\E \left[\ii \{ \zl \neq 0\}\zl\right]\right)^2, \nonumber \\
    &\leq \E \left[\left(\ii \{ \zl \neq 0\} \right)^2\right]\E \left[\left(\zl\right)^2\right], \label{eq:acschwarz}\\
 &= \E \left[\ii \{ \zl \neq 0\} \right]\E \left[\left(\zl\right)^2\right], \nonumber\\
    &= \pr[\zl \neq 0]\E\left[\left(\zl\right)^2\right]\nonumber,
\end{align}
where (\ref{eq:acschwarz}) is a consequence of the Cauchy-Schwarz inequality.
Thus, if $\E\left[\left(\zl\right)^2\right] \neq 0$ then we have
\begin{align}
    \pr[\zl \neq 0]\geq \frac{\left(\E\left[\zl\right]\right)^2}{\E\left[\left(\zl \right)^2\right]}. \label{eq:azerolawconditionprob}
\end{align}
From (\ref{eq:azerolawconditionprob}), note that if we can show that
\begin{align}
{\underset{n \rightarrow \infty}\liminf}
    \frac{\left(\E\left[\zl\right]\right)^2}{\E\left[\left(\zl \right)^2\right]} \geq 1, \label{eq:aliminf}
\end{align} then we get that $\limit\pr[\zl \neq 0]=1 $ 
 or equivalently $\limit \pr [\exists v \in \nodes : \deg(v)=k-1, t_v=1 ]=
1$. We then get the zero law by noting that $\pr[\exists v \in \nodes : \deg(v) \in \{0,\dots, k-1\}] \geq  \pr [\exists v \in \nodes : \deg(v)=k-1, t_v=1 ]$.

From exchangeability of indicator random variables $\ii\{\deg(v_i)=k-1,t_i=1\}$, $i=1,2,\dots,n$, we have 
\begin{align}
  \E \left [\left(\zl \right)^2 \right]  &=  \E \left[\left(\sum_{i=1}^{n} \ii\{\deg(v_i)=k-1, t_i=1\} \right)^2 \right],\nonumber\\
  &= n \E \left[\left(\ii\{\deg(v_1)=k-1, t_1=1\} \right)^2 \right]\nonumber\\ 
  &\quad+n (n-1) \E \left[ \ii\{\deg(v_1)=k-1, t_1=1\}\ii\{\deg(v_2)=k-1, t_2=1\}\right],\\
   &= n \pr \left[\deg(v_1)=k-1, t_1=1  \right]\nonumber\\ 
  &\quad+n (n-1) \pr \left[ \deg(v_1)=k-1, \deg(v_2)=k-1,t_1=1, t_2=1\right]
  . \label{eq:azero1}
\end{align}
Furthermore, the exchangeability of the indicator random variables $\ii\{\deg(v_i)=d\}$ also gives 
\begin{align}
   \E \left [\zl \right ]= n \pr \left[\deg(v_1)=k-1, t_1=1  \right].  \label{eq:azero2}
\end{align}
Combining (\ref{eq:azero1}) and (\ref{eq:azero2}), we get
\begin{align}
\hspace{-2mm}\frac{\E[\zl^2]}{(\E\left[\zl\right])^2}\hspace{-.5mm}=\hspace{-1mm} \frac{1}{\E\left[\zl\right]} \hspace{-.5mm}+\hspace{-.5mm}\frac{n-1}{n}
\dfrac{\pr[\deg(v_1)=\deg(v_2)=d, t_1=t_2=1]}{\left(\pr \left[\deg(v_1)=k-1, t_1=1  \right] \right)^2}.
\label{eq:aratio}
\end{align}
In view of (\ref{eq:azerolawconditionprob})--(\ref{eq:aratio}), we get $\pr[\zl \neq 0] \to 1$ if
the following two propositions are established.
  \begin{proposition}[Establishing zero-law in Theorem~\ref{theorem:th1}:~First moment result]
  {\sl 
  Consider a scaling $K_n:\N \rightarrow \N$, $0<\mu<1$ with $\kk = \mu+(1-\mu)K_n$, a positive integer $k\geq2$ and sequence $\g_n$ defined through (\ref{eq:hyp}).
     \label{prop:azero-lawfirstm}
  If $\g_n \rightarrow -\infty$, then the expected number of type-1 nodes in $\hh$ with degree $k-1$ approaches $\infty$, i.e., 
  \begin{align}
     \limit \E \left[\zl \right] = \infty \text{\qquad if \qquad} \g_n \rightarrow -\infty. \nonumber
  \end{align}
  }
  
  \end{proposition}
    \begin{proposition}[Establishing zero-law in Theorem~\ref{theorem:th1}:~Second moment result]
  {\sl 
  Consider a scaling $K_n:\N \rightarrow \N$, $0<\mu<1$ with $\kk = \mu+(1-\mu)K_n$, a positive integer $k\geq2$ and sequence $\g_n$ defined through (\ref{eq:hyp}).
    \label{prop:azero-lawsecondm}
  If $\g_n \rightarrow -\infty$, then
  \begin{align}
{\underset{n \rightarrow \infty}\limsup} \dfrac{\pr[\deg(v_1)=\deg(v_2)=k-1, t_1=t_2=1]}{\left(\pr \left[\deg(v_1)=k-1, t_1=1  \right] \right)^2} \leq 1 \nonumber.
  \end{align}
  }
  \end{proposition}
  Next, using Proposition \ref{prop:azero-lawfirstm} and Proposition \ref{prop:azero-lawsecondm} in (\ref{eq:aratio}) we get that if $\g_n \rightarrow -\infty$ then
  \begin{align}
{\underset{n \rightarrow \infty}\limsup} \dfrac{\E[\zl^2]}{(\E\left[\zl\right])^2} \leq 1 \nonumber.
  \end{align}
  which in turn yields (\ref{eq:aliminf}).
 Combining the fact that the ratio $ \dfrac{(\E\left[\zl\right])^2} {\E[\zl^2]}\leq 1$ with (\ref{eq:aliminf}) we get
    \begin{align}
{\limit} \dfrac{(\E\left[\zl\right])^2} {\E[\zl^2]}=1. \label{eq:alimeq}
  \end{align}
  Plugging (\ref{eq:alimeq}) into (\ref{eq:azerolawconditionprob}) we get that $\pr[\zl \neq 0]=1$. Finally, noting that $\pr[\exists v \in \nodes : \deg(v) \in \{0,\dots, k-1\}] \geq  \pr [\zl \neq 0]$, we get the zero law in Theorem~\ref{theorem:th1}.
$\myendpf$
\subsection{Useful decompositions}
We enumerate some mathematical statements used in our proof. 
\begin{enumerate}[]
    \item[1.] 
For any $x \in [0,1)$, it can be verified that
\begin{align}
    \log(1-x)&=-\int_{0}^{x} \dfrac{1}{1-t} dt=-x-\Psi(x), \label{eq:adeczero}
    \end{align}
    {where}
    \begin{align}
    \Psi(x)&:= \int_{0}^{x} \dfrac{t}{1-t} dt,~~0\leq x <1. \nonumber
\end{align}
Noting that $\Psi(x)$ is non-negative, (\ref{eq:adeczero}) gives the bound that for $x \in [0,1)$, we have
\begin{align}
    1-x \leq e^{-x}. \label{eq:adecone}
\end{align}
Furthermore, using L'Hospital's rule we get
\begin{align}{\underset{x \rightarrow 0}\lim} \dfrac{\Psi(x)}{x^2}=\dfrac{1}{2}.\label{eq:aLH}\end{align}
    \item[2.]
  If $x$ and $y$ are functions of $n$ such that $x = \oo{(1)}$ and $x^2y=\oo{(1)}$, then
  \begin{align}
        (1-x)^y=e^{-xy}(1+\oo{(1)}). \label{eq:ajun}
  \end{align}
 For a proof of this, see \cite[Fact 3]{Jun/K-Connectivity}.
\end{enumerate}
\section{Proof of Proposition \ref{prop:aone-law} (establishing one-law in Theorem~\ref{theorem:th1})}
\label{sec:aonelaw}
Here, we consider the case where $\limit \g_n = + \infty$. Recall that $\xd$ denotes the number of nodes with degree $d$ where $d \in \{1,2...,k-1\}$. 
We have
\begin{align}
\E\left[\xd\right]\nonumber&=\E\left[\sum_{i=1}^{n}\nonumber \ii\{\deg(v_i)=d\}\right],\nonumber\\
&= n \E[ \ii\{\deg(v_1)=d\}],\nonumber\\
&= n \pr[\dd(v_1)=d],\nonumber\\
&= n (\mu \pr[\dd(v_1)=d ~|~t_1=1]+{\color{black}(1-\mu)\pr[\dd(v_1)=d ~|~ t_1=2])}, \label{eq:aprob}
\end{align}
Note that if $\limit \g_n = + \infty$, from the scaling condition (\ref{eq:hyp}) in Theorem~\ref{theorem:th1}, it is evident that $\limit \kk= + \infty$ and thus $\forall k \in \N$, $\exists n_0$ such that $\forall n>n_0$, $\kk > k$. Consequently, for sufficiently large $n$, a type-2 node can never have degree less than $k$ for any finite value of $k$ and thus the second term in (\ref{eq:aprob}) vanishes. Thus
\begin{align}
\E\left[\xd\right]\nonumber&= n \mu \pr[\dd(v_1)=d~|~t_1=1]+{\color{black}0},\nonumber\\
&= n \mu {n-2 \choose d-1} \left(\dfrac{\kk}{n-1}\right)^{d-1}\left(1-\dfrac{\kk}{n-1}\right)^{n-1-d}, \label{eq:degcount}\\
&= n {\mu} \dfrac{(n-2)\dots ((n-2)-(d-2))}{(d-1)!}\left(\dfrac{\kk}{n-1}\right)^{d-1}\left(1-\dfrac{\kk}{n-1}\right)^{n-1-d},\nonumber\\
&= \dfrac{\mu}{(d-1)!} \cdot n \cdot {\dfrac{(n-2)}{n-1}\dots \dfrac{((n-2)-(d-2))}{n-1}} \kk^{d-1}\left(1-\dfrac{\kk}{n-1}\right)^{n-1-d},\nonumber\\
&= \dfrac{\mu}{(d-1)!}  \cdot n \cdot  \left(1-\dfrac{1}{n-1}\right)\dots \left(1-\dfrac{d-1}{n-1}\right)\kk^{d-1}\left(1-\dfrac{\kk}{n-1}\right)^{n-1-d},\label{eq:lemsim1}\\
&\leq \dfrac{\mu}{(d-1)!}  \cdot n \cdot  \left(1-\dfrac{1}{n-1}\right)\dots \left(1-\dfrac{d-1}{n-1}\right)\kk^{d-1}\exp{\left(-\dfrac{\kk(n-1-d)}{n-1}\right)},\label{eq:expineq}\\
&= \dfrac{\mu}{(d-1)!}  \cdot n \cdot  \left(1-\dfrac{1}{n-1}\right)\dots \left(1-\dfrac{d-1}{n-1}\right)\kk^{d-1}\exp{\left(-\kk\left(1-\dfrac{d}{n-1}\right)\right)}\nonumber,\\
&= \dfrac{\mu}{(d-1)!}   \left(1-\dfrac{1}{n-1}\right)\dots \left(1-\dfrac{d-1}{n-1}\right)\exp{ \left(\log n + (d-1) \log \kk-\kk\left(1-\dfrac{d}{n-1}\right)\right)}. \label{eq:ac1t1}
\end{align}
Here, (\ref{eq:expineq}) follows from the inequality $1-x\leq \exp(-x)~ \forall x \in [0,1)$; see (\ref{eq:adecone}). Next, we simplify the argument of the exponent in the right hand side of (\ref{eq:ac1t1}). We do so by substituting $\kk$ in terms of $\g_n$ using (\ref{eq:hyp}) and obtain
\begin{align}
& \log n + (d-1) \log \kk-\kk\left(1-\dfrac{d}{n-1}\right)\nonumber\\
&=\log n + (d-1) \log (\log n +(k-2)\log \log n+\g_n)-(\log n +(k-2)\log \log n+\g_n)\left(1-\dfrac{d}{n-1}\right),\nonumber\\
&=\log n + (d-1) \log \left(\log n \left(1+\dfrac{(k-2)\log \log n}{\log n}+\dfrac{\g_n}{\log n}\right)\right)-(\log n +(k-2)\log \log n+\g_n)\left(1-\dfrac{d}{n-1}\right),\nonumber\\
&=\log n + (d-1) \log \log n +(d-1)\log \left(1+\dfrac{(k-2)\log \log n}{\log n}+\dfrac{\g_n}{\log n}\right)\nonumber -(\log n +(k-2)\log \log n+\g_n)\left(1-\dfrac{d}{n-1}\right),\nonumber\\
&=-(k-1-d) \log \log n +(d-1)\log \left(1+\dfrac{(k-2)\log \log n}{\log n}+\dfrac{\g_n}{\log n}\right)\nonumber -\g_n\left(1-\dfrac{d}{n-1}\right)+d \dfrac{\log n}{n-1}+d(k-2)\dfrac{\log \log n}{n-1} ,\nonumber\\
&=-(k-1-d) \log \log n +(d-1)\log \left(1+\oo(1)+\dfrac{\g_n}{\log n}\right) -\g_n\left(1-\oo(1)\right)+\oo(1). \label{eq:azerolaw10}
\end{align}
Observe that for $n$ sufficiently large, 
$1+\oo(1)+\dfrac{\g_n}{\log n} \leq \g_n$ since $\g_n \rightarrow \infty$. 
Combining this with (\ref{eq:azerolaw10}) gives
\begin{align}
& \log n + (d-1) \log \kk-\kk\left(1-\dfrac{d}{n-1}\right)\nonumber\\
&\leq-(k-1-d) \log \log n +(d-1)\log \g_n -{\g_n}{(1-\oo{(1)})}+\oo(1),\nonumber\\
&=-(k-1-d) \log \log n -\g_n(1-\oo(1))+\oo(1). \label{eq:aexparg}
\end{align}
Combining (\ref{eq:ac1t1}) and (\ref{eq:aexparg}) we get
\begin{align}
\E\left[\xd\right]&\leq \dfrac{\mu}{(d-1)!} (1+\oo(1))\exp{ \left(\log n + (d-1) \log \kk-\kk\left(1-\dfrac{d}{n-1}\right)\right)}, \nonumber\\ 
&\leq \dfrac{\mu}{(d-1)!} (1+\oo(1))\exp{ \left(-(k-1-d) \log \log n -\g_n(1-\oo(1))+\oo(1)\right)} \label{eq:aonelawcomb}.
\end{align}
 Recall that $\xd$ is a non-negative random variable. Thus, from (\ref{eq:aonelawcomb}) we see that when $ \g_n \rightarrow \infty$, we have
$\limit \E\left[\xd\right] = 0$ for each $d=0,1,\dots,k-1$. This completes the proof of Proposition~\ref{prop:aone-law}.
$\myendpf$

\section{Proof of Proposition~\ref{prop:azero-lawfirstm}: (Establishing first moment result for zero-law in Theorem~\ref{theorem:th1})}
\label{sec:azerolawfirst}
Recall that $\zl$ denotes the number of type-1 nodes with degree $k-1$. In the succeeding arguments we show that if the sequence $\g_n\rightarrow -\infty$, then $\E \left[ \zl\right] \rightarrow + \infty$. From the exchangeability of 
and further conditioning on the event that the node is of type-1, we get
\begin{align}
\E\left[\zl\right]&=\E\left[\sum_{l=1}^{n} \ii\{\deg(v_i)=k-1, t_i=1\}\right],\\
&=n \pr[\deg(v_1)=k-1, t_1=1],\nonumber\\
&=n \pr[\deg(v_1)=k-1 ~| ~t_1=1]\pr[t_1=1],\nonumber\\
&= n \mu {n-2 \choose k-2} \left(\dfrac{\kk}{n-1}\right)^{k-2}\left(1-\dfrac{\kk}{n-1}\right)^{n-k}. \label{eq:degcount2}
\end{align}
\par Next, we inspect the term $\left(1-\frac{\kk}{n-1}\right)^{n-k}$ in Equation (\ref{eq:degcount2}). From scaling condition (\ref{eq:hyp}), it is evident that 
\begin{align}
\kk=\OO(\log n) \text{~~if~~} \limit \g_n = -\infty. \label{eq:aologn}
\end{align}
Using the decomposition given in (\ref{eq:adeczero}) 
with $x=\frac{\kk}{n-1}$ we get
\begin{align}
\left(1-\dfrac{\kk}{n-1}\right)^{n-k}&=  \exp\left(-\dfrac{\kk}{n-1}-\Psi\left(\dfrac{\kk}{n-1}\right)\right)^{n-k}, \nonumber\\
&=  \exp\left(-\left(\frac{n-k}{n-1}\right)\kk-{(n-k)}\Psi\left(\dfrac{\kk}{n-1}\right)\right) ,\nonumber\\
&=  \exp\left(-\left(1-\dfrac{k-1}{n-1}\right)\kk-{(n-k)}\Psi\left(\dfrac{\kk}{n-1}\right)\right) ,\nonumber\\
&= \exp\left(-\kk+(k-1)\dfrac{\kk}{n-1}-{(n-k)}\cdot \left(\dfrac{\kk}{n-1}\right)^2 \cdot \dfrac{\Psi\left(\dfrac{\kk}{n-1}\right)}{\left(\dfrac{\kk}{n-1}\right)^2}\right),\nonumber\\
&= \exp\left(-\kk\right)\exp\left((k-1)\dfrac{\kk}{n-1}\right)\exp\left(\dfrac{n-k}{n-1}\cdot \dfrac{\kk^2}{n-1} \cdot \dfrac{\Psi\left(\dfrac{\kk}{n-1}\right)}{\left(\dfrac{\kk}{n-1}\right)^2}   \right)\label{eq:ainterzerofirst1}.
\end{align}
Using (\ref{eq:aologn}) and (\ref{eq:aLH}) in (\ref{eq:ainterzerofirst1}) we get that
\begin{align}
\left(1-\dfrac{\kk}{n-1}\right)^{n-k}
&= \exp{\left(-\kk\right)}(1+\oo({1}))\label{eq:c2t2}.
\end{align}
Observe that (\ref{eq:degcount2}) is analogous to (\ref{eq:degcount}) evaluated at $d=k-1$. As done previously in \ref{sec:azerolawfirst}, combining (\ref{eq:c2t2}) and (\ref{eq:degcount2}), we get
\begin{align}
\E\left[\zl\right]&=(1+\oo(1)) \dfrac{\mu}{(k-2)!}  \exp(\log n + (k-2)\log \kk -\kk) \label{eq:z1}, \\
&= (1+\oo(1)) \dfrac{\mu}{(k-2)!}  \exp\left((k-2)\log \left(1 +\dfrac{(k-2) \log \log n }{\log n}+ \dfrac{\g_n}{\log n}\right) - \g_n \right).  \label{eq:z2}
\end{align}
Here, (\ref{eq:z2}) follows upon using the scaling condition (\ref{eq:hyp}).
Next, we showed  that $\limit\E\left[\zl\right]=+\infty$. Note that for any $n \in \N$, exactly one of the following statements is true.
\begin{enumerate}
\item[\bf{(i)}] $\g_n \leq -\dfrac{\log n}{2}$,
\item[\bf{(ii)}] $\g_n > -\dfrac{\log n}{2}$.
\end{enumerate}

\textbf{Case (i)}
For this case, $\g_n \leq -\dfrac{\log n}{2}$ and thus $\kk \leq \dfrac{\log n}{2}+(k-2)\log\log n$. Also, note that $\kk-1=(1-\mu)(K_n-1)>0$ and thus $\log \kk>0$. 
Using (\ref{eq:z1}) this yields
\begin{align}
\exp(\log n + (k-2)\log \kk -\kk) &\geq \exp(\log n -\kk) \label{eq:t1} ,\\
&\geq \exp( \log n-\dfrac{\log n}{2}-(k-2)\log\log n) ,\nonumber\\
&=  \exp\left( \dfrac{\log n}{2}-(k-2)\log\log n\right).\label{eq:ii}
\end{align}

\textbf{Case (ii)}
For this case, $\g_n > -\dfrac{\log n}{2}$ and thus $\dfrac{\g_n}{\log n}>\dfrac{-1}{2}$. We can now lower bound $\E\left[\zl\right]$ using (\ref{eq:z2}) in the following manner.
\begin{align}
 \exp\left(\log \left(1 +\dfrac{(k-2) \log \log n }{\log n}+ \dfrac{\g_n}{\log n}\right) - \g_n \right)
&>  \exp\left((k-2)\log \left(\dfrac{(k-2) \log \log n }{\log n}+ \dfrac{1}{2}\right) - \g_n \right),\nonumber\\
&\geq  \exp\left((k-2)\log \left( \dfrac{1}{2}\right) - \g_n \right),\nonumber\\
&= \left(\dfrac{1}{2}\right)^{k-2} \exp\left( - \g_n \right). \label{eq:i}
\end{align}

\noindent From (\ref{eq:z1}), (\ref{eq:z2}), (\ref{eq:i}) and (\ref{eq:ii}), it follows that for all $n$,
\begin{align}
\E\left[\zl\right]&\geq (1+\oo(1)) \dfrac{\mu}{(k-2)!} \min\left\{ \left(\dfrac{1}{2}\right)^{k-2} \exp\left(- \g_n \right),  \exp\left( \dfrac{\log n}{2}-(k-2)\log\log n\right)\right\},\label{eq:min}
\end{align}
From (\ref{eq:min}) we see that when $\limit \g_n = -\infty$, we have that $\limit \E\left[\zl\right]=+ \infty$.~~$\myendpf$
\section{Proof of Proposition~\ref{prop:azero-lawsecondm}: (Establishing second moment result for zero-law in Theorem~\ref{theorem:th1})}
\label{sec:azerolawsecond}
\par Now that we have proved Proposition~\ref{prop:azero-lawfirstm}, in order to establish the zero law we need to prove Proposition~\ref{prop:azero-lawsecondm} which states that if $\g_n \rightarrow -\infty$ then 
\begin{align}
 {\underset{n \rightarrow \infty}\limsup}\dfrac{\pr[\deg(v_1)=\deg(v_2)=k-1~|~t_1=t_2=1]}{{\left(\pr[\deg(v_1)=k-1~|~t_1=1]\right)^2}} \leq 1. 
\end{align}

\begin{figure}[h]
\centering\includegraphics[scale=0.29]{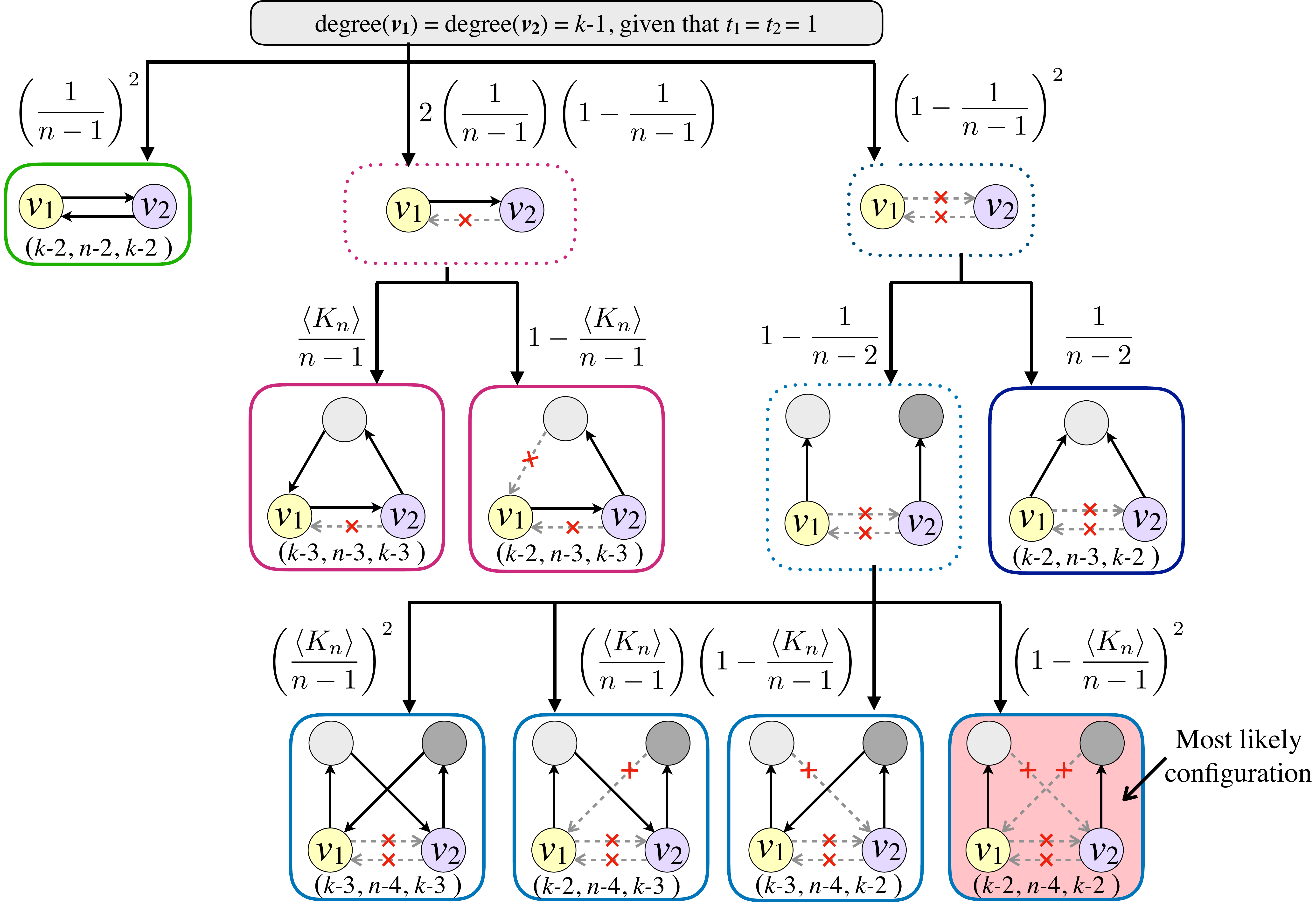}
\caption{ \sl Configurations for which nodes~$v_1$ and $v_2$ achieve the desired node degree of $k-1$ given that both these nodes are type-1. The weight given for a branch $A\rightarrow B$ gives the conditional probability of arriving at the next configuration $B$ from the current configuration $A$. In the tuple $(\alpha, m, \beta)$, $\alpha$ (respectively, $\beta$) denotes the number of additional edges needed to be drawn from $m$ nodes to obtain a degree of $k-1$ for node~$v_1$ (respectively, node~$v_2$). There are a total of 10 distinct realizations which result in the node degrees of nodes $v_1$ and $v_2$ being $k-1$ given that they are type-1 nodes. In the figure we only include one of the symmetric cases $\pr[\deg(v_1)=\deg(v_2)=k-1 ~|~ 2 \con1, 1 \dcon 2,t_1=t_2=1]$ and $\pr[\deg(v_1)=\deg(v_2)=k-1 ~|~ 1 \con1, 2 \dcon 1, t_1=t_2=1]$. 
}\label{fig:2} 
\end{figure}

Recall that $\kk/n-1$ is the probability with which a node under consideration is connected with node $v_1$. Moreover, every type-1 node has at least degree 1 corresponding to the selection it makes. Therefore, in order to have degree of $k-1$, a type-1 node needs to be picked by $k-2$ nodes other than  itself and the node selected by node $v_1$. Thus, we get 
\begin{align}
\pr[\deg(v_1)=k-1~|~t_1=1]&=  {n-2 \choose k-2} \left(\dfrac{\kk}{n-1}\right)^{k-2}\left(1-\dfrac{\kk}{n-1}\right)^{n-k}.
\end{align}
Next, we compute the ${\pr[\deg(v_1)=\deg(v_2)=k-1~|~t_1=t_2=1]}$. Let $\mathcal{E}$ denote the event that $\deg(v_1)=\deg(v_2)=k-1$ where $k\geq2$. Let $i \rightarrow j$ denote the event that node~$v_i$ picks node~$v_j$ and let $i \dcon j$ denote its compliment, i.e., the event that node~$v_i$ does not pick node~$v_j$. Further, conditioning on the edges drawn by node~$v_1$ and node~$v_2$, we get different cases as illustrated in Figure~\ref{fig:2}. Here, the weight given for a branch $A \rightarrow B$ denotes the conditional probability of occurrence of configuration $B$ given configuration $A$.
 ~For example, the event that neither node $v_1$ nor node $v_2$ picks the other occurs with probability given that both $v_1$ and $v_2$ are of  type-1 occurs with probability $\left(1-\frac{1}{n-1}\right)^2$. Further, given the event that neither node~$v_1$ picks node~$v_2$ nor node~$v_2$ picks node~$v_1$ denoted by $1 \dcon2, 2 \dcon 1,t_1=t_2=1$, the probability that node~$v_1$ and node~$v_2$ pick the same node is $\frac{1}{n-2}$. 
 
 Next, depending on whether $v_1$ and $v_2$ select each other, $\pr[\deg(v_1)=\deg(v_2)=k-1|t_1=t_2=1]$ can be decomposed as follows.
\begin{align}
&\pr[\event|t_1=t_2=1] \nonumber\\
&=\pr[\event ~|~ 1 \con2, 2 \con 1, t_1=t_2=1]\pr[1 \con2, 2 \con 1~|~t_1=t_2=1]\nonumber\\
&\quad+\pr[\event ~|~ 1 \con2, 2 \dcon 1, t_1=t_2=1]\pr[1 \con2, 2 \dcon 1~|~t_1=t_2=1]\nonumber\\
&\quad+\pr[\event ~|~ 2 \con1, 1 \dcon 2, t_1=t_2=1]\pr[2 \con1, 1 \dcon 2~|~t_1=t_2=1]\nonumber\\
&\quad+\pr[\event ~|~ 1 \dcon2, 2 \dcon 1,t_1=t_2=1]\pr[1 \dcon2, 2 \dcon 1~|~t_1=t_2=1]\nonumber\\
&=\pr[\event ~|~ 1 \con2, 2 \con 1,t_1=t_2=1]\dfrac{1}{(n-1)^2}+2\pr[\event ~|~ 1 \con2, 2 \dcon 1,t_1=t_2=1]\dfrac{1}{n-1}\left(1-\dfrac{1}{n-1}\right)\nonumber\\
&\quad+\pr[\event ~|~ 1 \dcon2, 2 \dcon 1,t_1=t_2=1]\left(1-\dfrac{1}{n-1}\right)^2,
\end{align}
where the last step follows from the symmetrical nature of events 
$\pr[\event ~|~ 2 \con1, 1 \dcon 2,t_1=t_2=1]$ and $\pr[\event ~|~ 1 \con2, 2 \dcon 1, t_1=t_2=1]$. Depending on the selections made by nodes $v_1$ and $v_2$, there can be multiple realizations which result in a degree of $k-1$ for both nodes $v_1$ and $v_2$. Figure~\ref{fig:2} enumerates all such realizations and their corresponding probabilities of occurrence. Corresponding to each realization, we mark the tuple $(\alpha, m, \beta)$, in which $\alpha$ (respectively, $\beta$) denotes the number of additional edges needed to be drawn from $m$ nodes in order to get the resulting degree of $k-1$ for node~$v_1$ (respectively, node~$v_2$). 
\begin{figure}[h]
\centering\includegraphics[scale=0.33]{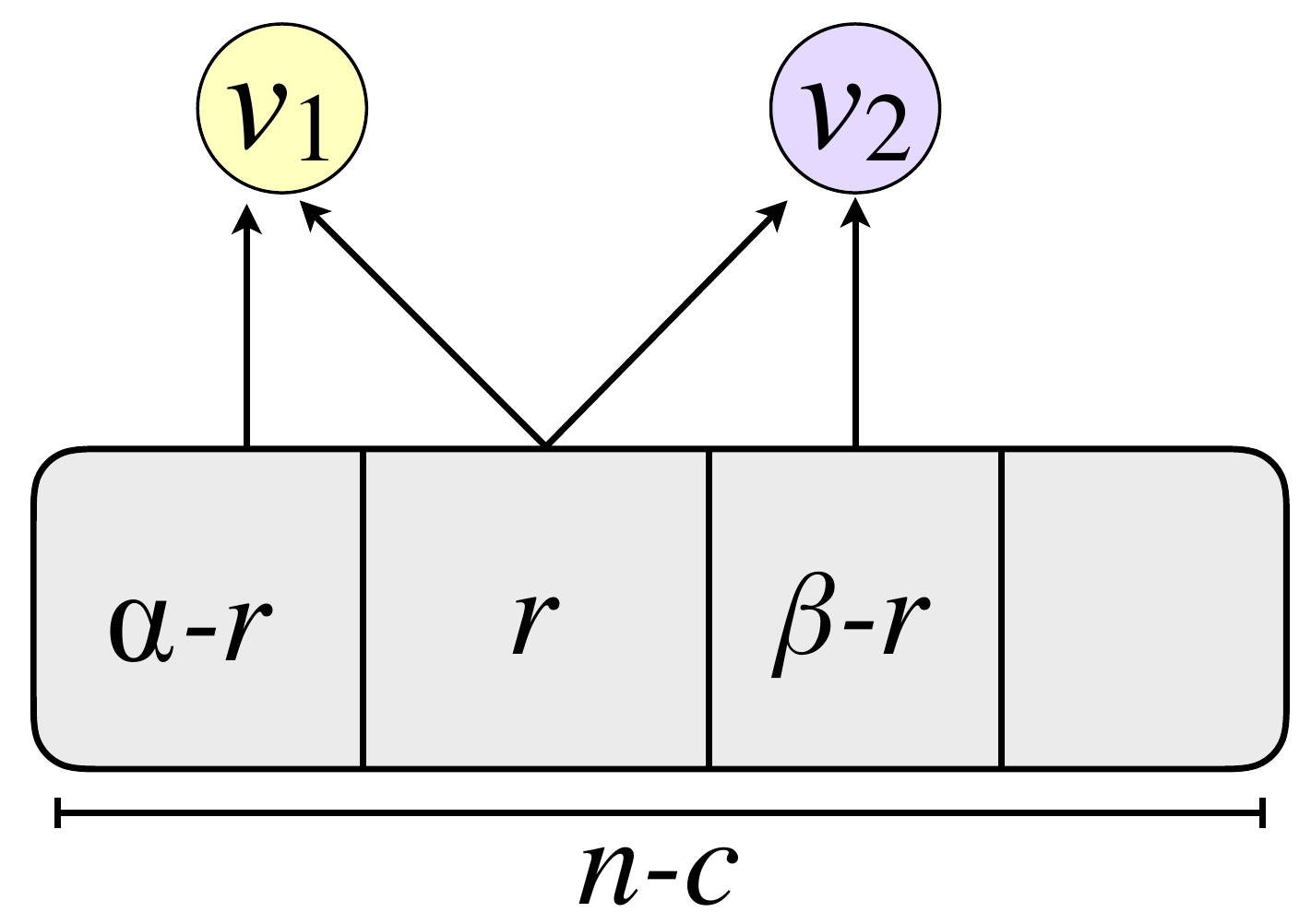}
\caption{\sl From a pool of $n-c$ nodes, $\alpha-r$ nodes draw an edge to node~$v_1$ but not node~$v_2$, $\beta-r$ nodes draw an outgoing edge to node~$v_2$ but not node~$v_1$, $r$ nodes draw an outgoing edge to both node~$v_1$ and node~$v_2$ and the remaining $n-c-\alpha-\beta+r$ nodes draw an edge to neither node~$v_1$ nor node~$v_2$.}\label{fig:3} 
\end{figure}

Let $B^{n-c}_{\alpha,\beta}$ denote the probability that out of a total of $n-c$ nodes, $\alpha$ nodes draw an edge to node~$v_1$ and $\beta$ nodes draw an outgoing edge to node~$v_2$.
 ~To evaluate $B^{n-c}_{\alpha,\beta}$, we need to consider different cases bases on the selections made by $v_1$ and $v_2$.
Recall that our goal is to show that
\begin{align}
    {\underset{n \rightarrow \infty}\limsup}&\dfrac{\pr[\deg(v_1)=\deg(v_2)=k-1|t_1=t_2=1]}{\left({\pr[\deg(v_1)=k-1~|~t_1=1]}\right)^2} \leq 1.
\end{align}
We can write $\pr[\event|t_1=t_2=1]$ by summing over different configurations shown in Figure~\ref{fig:2} as follows.
\begin{align}
\pr[\event|t_1=t_2=1] &=\dfrac{1}{(n-1)^2}B^{n-2}_{k-2,k-2} +\dfrac{2}{n-1}\left(1-\dfrac{1}{n-1}\right)B^{n-3}_{k-3,k-3}+\dfrac{2}{n-1}\left(1-\dfrac{1}{n-1}\right)B^{n-3}_{k-2,k-3}\nonumber\\
&\quad+\left(1-\dfrac{1}{n-1}\right)^2\dfrac{1}{n-2}B^{n-3}_{k-2,k-2}+\left(1-\dfrac{1}{n-1}\right)^2\left(1-\dfrac{1}{n-2}\right)\left(\dfrac{ \kk}{n-1} \right)^2 B^{n-4}_{k-3,k-3} \nonumber\\
&\quad+\left(1-\dfrac{1}{n-1}\right)^2\left(1-\dfrac{1}{n-2}\right)\cdot 2 \cdot \dfrac{ \kk}{n-1}\left(1-\dfrac{\kk}{n-1}\right)B^{n-4}_{k-2,k-3}\nonumber\\
&\quad+\left(1-\dfrac{1}{n-1}\right)^2\left(1-\dfrac{1}{n-2}\right)\left(1-\dfrac{\kk}{n-1}\right)^2 B^{n-4}_{k-2,k-2}. \label{eq:afullexpr}
\end{align}
In order to establish Proposition~\ref{prop:azero-lawsecondm}, we postulate the following lemma.
\begin{lemma}
\sl 
  Consider a scaling $K_n:\N \rightarrow \N$, $0<\mu<1$ with $\kk = \mu+(1-\mu)K_n$, a positive integer $k\geq2$ and sequence $\g_n$ defined through (\ref{eq:hyp}).
  If $\g_n \rightarrow -\infty$, then
$ \ \dfrac{B^{n-c}_{\alpha,\beta}}{\left({\pr[\deg(v_1)=k-1~|~t_1=1]}\right)^2}  = 
\dfrac{((k-2)!)^2}{\alpha! \beta!} \kk^{\alpha+\beta+4-2k}(1+\oo{(1)}).$
\label{lemma:lemlast}
\end{lemma}

\noindent Table~\ref{table:tab11} enumerates the limit that  $\frac{B^{n-c}_{\alpha,\beta}}{{\pr^2[\deg(v_1)=k-1~|~t_1=1]}}$ approaches as $n\rightarrow\infty$ for different pairs of $\alpha$ and $\beta$ arising in our analysis (Figure \ref{fig:2}).
\begin{table}[htbp]
\caption{Evaluation of the expression in Lemma~\ref{lemma:lemlast} for different $(\alpha,\beta)$ arising in Figure~\ref{fig:2}}
\begin{center}
 \begin{tabular}{|c| c| c| c|} 
 \hline
$\alpha$ & $\beta$ & $ \ \dfrac{B^{n-c}_{\alpha,\beta}}{\left({\pr[\deg(v_1)=k-1~|~t_1=1]}\right)^2} $ \\ [0.5ex] 
 \hline\hline
 $k-3$ &  $k-3$ & $ \dfrac{(k-2)^2}{\kk^2} (1+\oo(1))$ \\ 
 \hline
 $k-3$ &  $k-2$ & $ \dfrac{k-2}{\kk}(1+\oo(1))$  \\
 \hline
 $k-2$ &  $k-3$ & $ \dfrac{k-2}{\kk}(1+\oo(1))$  \\
 \hline
  $k-2$ &  $k-2$ & $1+\oo(1)$ \\
 \hline
 \end{tabular}
\label{table:tab11}
\end{center}
\end{table}
From Figure~\ref{fig:2} and Table~\ref{table:tab11}, it can be seen that as $n\rightarrow \infty$ the only non-zero term appearing in (\ref{eq:afullexpr}) corresponds to the case when neither node $v_1$ nor $v_2$ pick each other and $v_1$ and $v_2$ nodes select distinct nodes. Further, the node that is selected by $v_1$ (respectively, $v_2$) does not select $v_2$ (respectively, $v_1$).
As $n\rightarrow \infty$, this configuration has a probability of $1$ and is the dominant term in (\ref{eq:afullexpr}). Moreover, from Figure~\ref{fig:2} note that for this case $\alpha=\beta=k-2$ and using Lemma~\ref{lemma:lemlast}, we have $\limit \ \frac{B^{n-c}_{\alpha,\beta}}{\left({\pr[\deg(v_1)=k-1~|~t_1=1]}\right)^2} =1$ (Table \ref{table:tab11}). Therefore,$\limit\frac{\pr[\deg(v_1)=\deg(v_2)=k-1|t_1=t_2=1]}{\left({\pr[\deg(v_1)=k-1~|~t_1=1]}\right)^2} = 1$ and thus $  {\underset{n \rightarrow \infty}\limsup}\frac{\pr[\deg(v_1)=\deg(v_2)=k-1|t_1=t_2=1]}{\left({\pr[\deg(v_1)=k-1~|~t_1=1]}\right)^2} = 1$. Thus, Proposition~\ref{prop:azero-lawsecondm} holds as a consequence of Lemma~\ref{lemma:lemlast}. 
$\myendpf$
\section{Proof of Lemma~\ref{lemma:lemlast}}
We proceed to prove Lemma~\ref{lemma:lemlast} which states that if $\g_n \rightarrow - \infty$, then we have
$$ \ \dfrac{B^{n-c}_{\alpha,\beta}}{\left({\pr[\deg(v_1)=k-1~|~t_1=1]}\right)^2}  = 
\dfrac{((k-2)!)^2}{\alpha! \beta!} \kk^{\alpha+\beta+4-2k}(1+\oo{(1)}).$$
Recall that $B^{n-c}_{\alpha,\beta}$ denotes the probability that out of a pool of $n-c$ nodes, $\alpha$ nodes draw an edge to node~$v_1$ and $\beta$ nodes draw an outgoing edge to node~$v_2$. Here, note that it is possible that some type-2 nodes draw edges to both nodes~$v_1$ and $v_2$. In order to compute $B^{n-c}_{\alpha,\beta}$, we introduce $A^{n-c}_{\alpha-r,r,\beta-r}$ as the probability that out of a total of $n-c$ nodes, $\alpha-r$ nodes draw an edge to node~$v_1$ but not node~$v_2$, $\beta-r$ nodes draw an outgoing edge to node~$v_2$ but not node~$v_1$, $r$ nodes draw an outgoing edge to both node~$v_1$ and the remaining nodes draw an edge to neither (Figure~\ref{fig:3}).
 ~If node $v_1$ (respectively, $v_2$) needs ($\alpha$, respectively $\beta$) additional edges from the remaining $n-c$ to achieve a degree of $k-1$, then $\pr [\deg(v_1)=\deg(v_2)=k-1 ~|~ t_1=t_2=1]=B^{n-c}_{\alpha,\beta}$. We can express $B^{n-c}_{\alpha,\beta}$ in terms of $A^{n-c}_{\alpha-r,r,\beta-r}$ where $0\leq r\leq \min\{\alpha, \beta\}$ as follows.
\begin{align}
    B^{n-c}_{\alpha,\beta}&=\sum_{r=0}^{\min\{\alpha,\beta\}} A^{n-c}_{\alpha-r,r,\beta-r}. \nonumber
\end{align}
In order to explicitly compute, $A^{n-c}_{\alpha-r,r,\beta-r}$ we denote
\begin{align}
p_{12}&:=\pr[v_x\con v_1,v_x\con v_2],\nonumber\\
p_{\bar{1}2}&:=\pr[v_x\dcon v_1,v_x\con v_2],\nonumber\\
p_{1\bar{2}}&:=\pr[v_x\con v_1,v_x\dcon v_2],\nonumber\\
p_{\bar{1}\bar{2}}&:=\pr[v_x\dcon v_1,v_x\dcon v_2]\nonumber,
\end{align}
where node $v_x$ is a node from a pool of $n-c$ nodes (Figure \ref{fig:2}). 
Observe that
\begin{align}
A^{n-c}_{\alpha-r,r,\beta-r}&= {n-c \choose \alpha-r}{n-c-\alpha+r \choose r}{n-c-\alpha \choose \beta-r} (p_{1\bar{2}})^{\alpha-r}(p_{1{2}})^r(p_{\bar{1}{2}})^{\beta-r}(p_{\bar{1}\bar{2}})^{n-c-\alpha-\beta+r},\nonumber\\
&=\dfrac{(n-c)!}{(\alpha-r)! r!  (\beta-r)! (n-c-(\alpha+\beta)+r)!} (p_{1\bar{2}})^{\alpha+\beta-2r}(p_{1{2}})^r(p_{\bar{1}\bar{2}})^{n-c-(\alpha+\beta)+r}\label{eq:adef}.
\end{align}

In order to prove Lemma~\ref{lemma:lemlast}, we need the following intermediate results which show that $A^{n-c}_{\alpha-r,r,\beta-r}$ is a monotone decreasing sequence in $r$ with $A^{n-c}_{\alpha-(r+1),r+1,\beta-(r+1)}= \oo(1)A^{n-c}_{\alpha-r,r,\beta-r} $ for all $r=0,1,2,\dots,\min \{ \alpha, \beta\}$. Next, we compute the dominant term corresponding to $r=0$ in the finite sum $B^{n-c}_{\alpha,\beta}=\sum_{r=0}^{\min\{\alpha,\beta\}}$

\begin{lemma} \label{lemma:ratioterms}
\sl 
  Consider a scaling $K_n:\N \rightarrow \N$, $0<\mu<1$ with $\kk = \mu+(1-\mu)K_n$, a positive integer $k\geq2$ and sequence $\g_n$ defined through (\ref{eq:hyp}).
If $\g_n \rightarrow - \infty$, then
\begin{align}
 \dfrac{A^{n-c}_{\alpha-(r+1),(r+1),\beta-(r+1)}}{A^{n-c}_{\alpha-r,r,\beta-r}}
&=\oo{(1)}\text{\quad for all }r=0,1,\dots,\min \{ \alpha,\beta\}. \nonumber
\end{align}
\begin{lemma}\label{lemma:r0}
\sl 
  Consider a scaling $K_n:\N \rightarrow \N$, $0<\mu<1$ with $\kk = \mu+(1-\mu)K_n$, a positive integer $k\geq2$ and sequence $\g_n$ defined through (\ref{eq:hyp}).
If $\g_n \rightarrow - \infty$, then
\begin{align}
 A^{n-c}_{\alpha,0,\beta}
&= \ \dfrac{1}{\alpha! \beta !} \kk^{\alpha+\beta}\exp(-2\kk)(1+\oo{(1)}). \nonumber
\end{align}
\end{lemma}
\end{lemma}
Recall that we can express $B^{n-c}_{\alpha,\beta}$ in terms of $A^{n-c}_{\alpha-r,r,\beta-r}$ as follows. 
\begin{align}
    B^{n-c}_{\alpha,\beta}&=\sum_{r=0}^{\min\{\alpha,\beta\}} A^{n-c}_{\alpha-r,r,\beta-r}. \label{eq:astar}
\end{align}
Using Lemma~\ref{lemma:ratioterms} in (\ref{eq:astar}), we get
\begin{align}
    B^{n-c}_{\alpha,\beta}&= A^{n-c}_{\alpha,0,\beta}\left(1+\oo{(1)}+(\oo{(1))^2}+\dots+
    (\oo{(1)})^{\min\{\alpha,\beta\}}\right). \nonumber\\
    &=A^{n-c}_{\alpha,0,\beta}\left(1+\oo{(1)}\right), \label{eq:aoosum}
\end{align}
where (\ref{eq:aoosum}) follows from noting that $\alpha, \beta$ are finite constants and thus (\ref{eq:astar}) corresponds to a finite sum.
Using Lemma~\ref{lemma:r0} in (\ref{eq:aoosum}), we get
\begin{align}
 B^{n-c}_{\alpha,\beta}= \ \dfrac{1}{\alpha! \beta !} \kk^{\alpha+\beta}\exp(-2\kk)(1+\oo{(1)}). \label{eq:num}
\end{align}
Observe that 
\begin{align}
    \pr[\deg(v_1)=k-1~|~t_1=1]&={n-2 \choose k-2} \left(\dfrac{\kk}{n-1}\right)^{k-2}\left(1-\dfrac{\kk}{n-1}\right)^{n-k}. \nonumber
\end{align}
Arguing as before and using (\ref{eq:c2t2}), we get
\begin{align}
    \pr[\deg(v_1)=k-1~|~t_1=1]&=\dfrac{1}{(k-2)!} \kk^{k-2}\exp{(-\kk)}(1+\oo{(1)}) \label{eq:rootdenom}.
\end{align}
Combining (\ref{eq:rootdenom}) and (\ref{eq:num}), we get the desired result,
\begin{align}
    \dfrac{B^{n-c}_{\alpha,\beta}}{\left({\pr[\deg(v_1)=k-1~|~t_1=1]}\right)^2}  = 
\dfrac{(k-2)!^2}{\alpha! \beta!} \kk^{\alpha+\beta+4-2k}(1+\oo{(1)}). \nonumber
\end{align}
The proof of Lemma~\ref{lemma:lemlast} is now complete.
$\myendpf$
\noindent{\emph{E}.I. {~\sl Proof of Lemma~\ref{lemma:ratioterms}}}
\\
Next, we prove Lemma~\ref{lemma:ratioterms} which states that if $\g_n \rightarrow - \infty$, then
\begin{align}
 \dfrac{A^{n-c}_{\alpha-(r+1),(r+1),\beta-(r+1)}}{A^{n-c}_{\alpha-r,r,\beta-r}}
&=\oo{(1)}. \nonumber
\end{align}
From (\ref{eq:adef}), observe that
 \begin{align}
 \dfrac{A^{n-c}_{\alpha-(r+1),(r+1),\beta-(r+1)}}{A^{n-c}_{\alpha-r,r,\beta-r}}   &=\dfrac{(\alpha-r)(\beta-r)}{(r+1)(n-c-(\alpha+\beta)+r)}\cdot\dfrac{p_{12}p_{\bar{1}\bar{2}}}{(p_{1\bar{2}})^2}.\label{eq:ratio1}
\end{align}
Next, we compute $\limit\frac{p_{12}p_{\bar{1}\bar{2}}}{(p_{1\bar{2}})^2}$.
We ascertain the probablities $p_{12}$, $p_{\bar{1}2}$, $p_{1\bar{2}}$ and $p_{\bar{1}\bar{2}}$. Observe that a type-1 node can only select one amongst nodes $v_1$ or $v_2$. Thus, both nodes $v_1$ and $v_2$ can be selected by only type-2 nodes. This gives
\begin{align}
p_{12}&=\mu. 0+(1-\mu) \dfrac{{n-3 \choose \K-2}}{{n-1 \choose \K}},\nonumber\\
&=(1-\mu)\dfrac{(\K)(\K-1)}{(n-1)(n-2)}.\label{eq:ap12}
\end{align}
Similarly, we can obtain the probability that exactly one amongst $v_1$ or $v_2$ is picked and the probability that neither $v_1$ nor $v_2$ is picked by a third node. Moreover from symmetry, $p_{\bar{1}2}=p_{1\bar{2}}$. We have
\begin{align}
p_{\bar{1}2}=p_{1\bar{2}}&=\mu\dfrac{1}{n-1}+(1-\mu)\dfrac{{n-3 \choose \K-1}}{{n-1 \choose \K}} \nonumber\\
&=\mu\dfrac{1}{n-1}+(1-\mu)\dfrac{(n-K_n-1)(K_n)}{(n-1)(n-2)}, \nonumber\\
&=\mu \dfrac{1}{n-1}+ (1-\mu) \dfrac{\K}{n-1}-(1-\mu)\dfrac{\K}{n-1}\dfrac{\K-1}{n-2},\nonumber\\
&= \dfrac{\kk}{n-1} \left(1-(1-\mu)\dfrac{\K}{\kk} \dfrac{\K-1}{n-2} \right),\nonumber\\
&= \dfrac{\kk}{n-1}(1+\oo(1)) \label{eq:aproofpn12}
\end{align}
 where the last step follows from noting that $1 \leq \kk \leq \K$ and $\kk=\OO(\log n)$.
Next, we simplify $ p_{\bar{1}\bar{2}}$ as follows.
\begin{align}
{\color{black}p_{\bar{1}\bar{2}}}&=\mu \dfrac{n-3}{n-1}+(1-\mu) \dfrac{{n-3 \choose \K}}{{n-1 \choose \K}}, \nonumber\\
&=\mu\dfrac{n-3}{n-1}+(1-\mu)\dfrac{(n-K_n-1)(n-K_n-2)}{(n-1)(n-2)},\nonumber\\
    &=\mu(1-\dfrac{2}{n-1})+(1-\mu)(1-\dfrac{\K}{n-1})(1-\dfrac{\K}{n-2}),\nonumber\\
    &= 1-\left(\dfrac{2\mu}{n-1}+(1-\mu)\frac{\K}{n-1}+(1-\mu)\frac{\K}{n-2} \right)+(1-\mu)\frac{\K}{n-1}\frac{\K}{n-2},\nonumber\\
    &= 1-\left(\dfrac{2\mu}{n-1}+(1-\mu)\frac{\K}{n-1}+(1-\mu)\frac{\K}{n-1}\left(1+\frac{1}{n-2}\right) \right)+(1-\mu)\frac{\K}{n-1}\frac{\K}{n-2},\nonumber\\
     &= 1-2\left(\dfrac{\mu+(1-\mu)\K}{n-1}\right)+(1-\mu)\frac{\K}{n-1}\frac{\K-1}{n-2},\nonumber\\
     &= 1-\left( \dfrac{2\kk}{n-1}-(1-\mu)\frac{\K}{n-1}\frac{\K-1}{n-2}\right).\label{eq:fact2app} 
\end{align}
When $\g_n \rightarrow -\infty$ then $\kk$ scales as $\OO(\log n)$ and an implication of (\ref{eq:fact2app}) is that 
\begin{align}
{\color{black}p_{\bar{1}\bar{2}}}&=
1+ \oo(1). \label{eq:fact2appsimple}
\end{align}
Using (\ref{eq:ap12}), (\ref{eq:aproofpn12}) and (\ref{eq:fact2appsimple}),
\begin{align}
 \dfrac{(p_{1\bar{2}})^2}{p_{\bar{1}\bar{2}}p_{12}} \nonumber 
&=  \dfrac{\dfrac{\kk^2}{(n-1)^2}(1+\oo(1))^2}{(1+\oo(1))(1-\mu)\dfrac{\K}{n}\dfrac{(\K-1)}{n}},\nonumber\\
&= \dfrac{{\kk^2}(1+\oo(1))^2\dfrac{{n^2}}{(n-1)^2}}
{(1+\oo(1))(1-\mu)\K (\K-1)}\nonumber\\
&=  \dfrac{\left(\mu+(1-\mu),{K_n}\right)^2}
{(1-\mu)\K (\K-1)},\nonumber\\
&= \dfrac{\mu^2}{(1-\mu)K_n(K_n-1)}+\dfrac{(1-\mu)K_n}{K_n-1}+\dfrac{2 \mu}{K_n-1}=\Theta(1). \nonumber
\end{align}
Thus, 
\begin{align}
\dfrac{p_{\bar{1}\bar{2}}p_{12}}{(p_{1\bar{2}})^2}
\label{eq:ratio2}
&=\Theta(1).
\end{align}
Combining (\ref{eq:ratio1}) and (\ref{eq:ratio2}), we get the desired result. 
\begin{align}
 \dfrac{A^{n-c}_{\alpha-(r+1),(r+1),\beta-(r+1)}}{A^{n-c}_{\alpha-r,r,\beta-r}}
&=\oo{(1)}.
\end{align}

$ ~~\myendpf$
This completes the proof of Lemma~\ref{lemma:ratioterms}.\\ \\
\noindent{\emph{E}.II. {~\sl Proof of Lemma~\ref{lemma:r0}}}\\
We need to prove Lemma~\ref{lemma:r0} which states that if $\g_n \rightarrow - \infty$, then
\begin{align}
 A^{n-c}_{\alpha,0,\beta}
&= \ \dfrac{1}{\alpha! \beta !} \kk^{\alpha+\beta}\exp(-2\kk)(1+\oo{(1)}). \nonumber
\end{align}
Substituting $r=0$ in (\ref{eq:adef}), we get
    \begin{align}
    A^{n-c}_{\alpha,0,\beta}
    &=\dfrac{(n-c)!}{\alpha!  \beta! (n-c-(\alpha+\beta))!} (p_{1\bar{2}})^{\alpha+\beta}(p_{\bar{1}\bar{2}})^{n-c-(\alpha+\beta)} \label{eq:azero00}
    \end{align}
Before proceeding with the proof of Lemma~\ref{lemma:r0}, we first show that when $\g_n \rightarrow - \infty$,
 \begin{align}
 (p_{\bar{1}\bar{2}})^{n-u}
 &= \exp{(-2\kk)}(1+\oo{(1)}), \label{eq:apnlarge2kk}
 \end{align}

Consider $u \in \mathbb{N}$, $1 \leq u < n$. Using (\ref{eq:ajun}) in (\ref{eq:fact2app}) with $y=n-u$ and $x=\frac{2\kk}{n-1}-(1-\mu)\frac{\K}{n-1}\frac{\K-1}{n-2}$ and noting that $x^2y=\oo(1)$, we get
  
\begin{align}
 (p_{\bar{1}\bar{2}})^{n-u} &\sim \exp{\left(-\dfrac{2\kk}{n-1}+(1-\mu)\frac{\K}{n-1}\frac{\K-1}{n-2}\right)(n-u)},\nonumber\\
&= \exp{(-2\kk)}(1+\oo{(1)}). \nonumber
\end{align}
 Finally, substituting $p_{1\bar{2}}$ and $p_{\bar{1}\bar{2}}$ in (\ref{eq:azero00}) using (\ref{eq:aproofpn12}) and (\ref{eq:apnlarge2kk}) combined with the fact that $\kk=\OO(\log n)$ gives
    \begin{align}
     A^{n-c}_{\alpha,0,\beta}
    &= \ \dfrac{(n-c)!}{\alpha!  \beta! (n-c-(\alpha+\beta))!} (p_{1\bar{2}})^{\alpha+\beta}(p_{\bar{1}\bar{2}})^{n-(c+\alpha+\beta)},\nonumber\\
    &=  \ \dfrac{(n-c)!}{\alpha!  \beta! (n-c-(\alpha+\beta))!} \left(\dfrac{\kk}{n-1}(1+\oo(1))\right)^{\alpha+\beta}\exp{(-2\kk)}(1+\oo{(1)}),\nonumber\\
    &= \ \dfrac{1}{\alpha ! \beta !} \kk^{\alpha+\beta} \exp{(-2 \kk)} \dfrac{\Pi_{i=0}^{\alpha+\beta-1}(n-c-i)}{(n-1)^{\alpha+\beta}}(1+\oo{(1)}),\nonumber\\
    &= \ \dfrac{1}{\alpha! \beta !} \kk^{\alpha+\beta}\exp(-2\kk)(1+\oo{(1)}),
    \end{align} 
    which completes the proof for Lemma~\ref{lemma:r0}.
    \myendpf




\section{Proof of Lemma~\ref{lemma:simplified}} 
If $|\g_n|=\oo{(\log n)}$, then using (\ref{eq:hyp}), we see that $\kk= \Theta (\log n)$. Consequently, as $n \rightarrow \infty$, type-2 nodes cannot have a finite degree. Observe that with $\kk= \Theta (\log n)$, (\ref{eq:c2t2}) holds true. (\ref{eq:c2t2}) can then be substituted in (\ref{eq:lemsim1}) and simplified as done previously in (\ref{eq:azerolaw10}) to get that the expected number of nodes with degree $d$ satisfies
\begin{align*}
\mathbb{E}\left[Z_{n,d}\right] = \Theta(1) \exp\left\{-(k-1-d)\log \log n - \gamma_n\right\}.
\end{align*}
\myendpf
\end{document}